\documentclass[twocolumn,aps,prb,showpacs,superscriptaddress,amsfonts,amssymb,amsmath]{revtex4}
\usepackage{amsmath}
\usepackage{graphicx}

\begin{document}
\title{The effect of inter-Landau-band mixing on extended states in quantum Hall system}
\author{Gang Xiong}
\affiliation{Physics Department, The Hong Kong
 University of Science and Technology, Clear Water Bay,
Hong Kong SAR, China} \affiliation{International Center for
Quantum Structures, Institute of Physics, Chinese Academy of
Sciences , Beijing 100080, P. R. China}
\affiliation{Physics
Department, Wuhan University, Wuhan 430072, P. R. China}
\author{Shi-Dong Wang} \affiliation{Physics Department, The
Hong Kong
 University of Science and Technology, Clear Water Bay,
Hong Kong SAR, China}
\author{Qian Niu}
\affiliation{Physics Department, The University of Texas at
 Austin, Austin, Texas 78712-1081}
\affiliation{International Center for Quantum Structures,
Institute of Physics, Chinese Academy of Sciences , Beijing
100080, P. R. China}
\author{Yupeng Wang}
\affiliation{International Center for Quantum Structures,
Institute of Physics, Chinese Academy of Sciences , Beijing
100080, P. R. China}
\author{X. C. Xie}
\affiliation{Physics Department, Oklahoma State University ,
Stillwater, OK 74078}
\affiliation{International Center for Quantum Structures,
Institute of Physics, Chinese Academy of Sciences , Beijing
100080, P. R. China}
\author{De-Cheng Tian}
\affiliation{Physics Department, Wuhan
University, Wuhan 430072, P. R. China}
\author{X. R. Wang}
\affiliation{Physics Department, The Hong Kong
 University of Science and Technology, Clear Water Bay,
Hong Kong SAR, China}

\date{Draft on \today}

\begin{abstract}
The effect of inter-Landau-band mixing on electron localization in
an integer quantum Hall system is studied. We find that mixing of
localized states with {\it opposite chirality} tends to delocalize
the states. This delocalization effect survives the quantum
treatment. Extended states form bands because of this mixing, as
we show through a numerical calculation on a two-channel network
model. Based on this result, we propose a new phase diagram with a
narrow {\it metallic} phase that separates any neighboring QH
phases from each other and also separates each of them from the
insulating phase. We reanalyzed the data from recent non-scaling
experiments, and show that they are {\it consistent} with our
theory.
\end{abstract}

 \maketitle

\section{Introduction}
According to the scaling theory of localization\cite {abrahams},
all electrons in a disordered two-dimensional system are localized
in the absence of a magnetic field. In the presence of a strong
magnetic field $B$, a series of disorder-broadened Landau bands
(LBs) will appear, and an extended state resides at the center of
each band while states at other energies are
localized\cite{pruisken}. The integrally quantized Hall plateaus
(IQHP) are observed when the Fermi level lies in localized states,
with the value of the Hall conductance, $\sigma_{xy}=ne^2/h$,
related to the number of occupied extended states($n$). Many
previous studies\cite{wei,kivelson,dzliu,wang,yang,
galstyan,sheng1,sheng2,haldane,jiang,tkwang,glozman,kravchenko,
song,xrw,hilke} have been focused on so-called plateau
transitions. The issue there is how the Hall conductance jumps
from one quantized value to another when the Fermi level crosses
an extended state. There are two competing proposals. One is the
global phase diagram\cite{kivelson} based on the levitation of
extended states conjectured by Khmelnitskii\cite{khme} and
Laughlin\cite{laughlin}. A crucial prediction of this phase
diagram is that an integer quantum Hall effect (IQHE) state $n$ in
general can only go into another IQHE states $n\pm 1$, and that a
transition into an insulating state is allowed only from the $n=1$
state. The other is so-called direct transition phase
diagram\cite{dzliu} in which transitions from any IQHE state to
the insulating phase are allowed when the disorder is increased at
fixed $B$. So far, most experiments\cite{kravchenko,song} are
consistent with the direct transition phase diagram although the
early experiments were interpreted in terms of the global phase
diagram.

One important yet overlooked issue regarding IQHE is the {\it
nature} of both plateau-plateau and plateau-insulator transitions.
In all existing theoretical studies, these transitions are assumed
to be continuous quantum phase transitions. This assumption is
mainly due to the early scaling experiments\cite{wei}. The
fingerprint of a continuous phase transition is scaling laws
around the transition point. In the case of IQHE, it means
algebraic divergence of the slope of the longitudinal resistance
in temperature $T$ at the transition point. However, recent
experiments\cite{hilke} showed that such slopes remain finite when
the curves are extrapolated to $T=0$. This implies a {\it
non-scaling} behavior around a transition point, contradicting
the expectation of continuous quantum phase transitions suggested
by the theories. Thus the nature of these transitions should be
re-examined.

The samples used in the non-scaling experiments\cite{hilke} are
relatively dirty, and strong disorders should lead to a strong
inter-Landau-band mixing. In a recent letter\cite {xiong}, we
showed that the single extended state at each LB center broadens
into a narrow band of extended states when the interband mixing of
{\it opposite chirality} is taken into account. A narrow metallic
phase exists between two adjacent IQHE phases and between an IQHE
phase and the insulating phase. A plateau-plateau transition
corresponds to two consecutive quantum phase transitions instead
of one as suggested by existing theories. In this paper we shall
present the detailed description of this study.

The paper is organized as follows. The semiclassical network model
for two coupled LBs is illustrated in Sec. II. It is shown that
mixing of localized states of {\it opposite chirality} tends to
delocalize a state while mixing of states of the same chirality
does not. Our approach, level-statistics technique, is described
in Sec. III. In Sec. IV our numerical results and discussions are
presented, and the original data from the non-scaling experiments
are reanalyzed according to two quantum phase transition points in
each IQHP-insulator transition. The conclusions of this paper are
summarized in Sec. V.

\section{The semiclassical model including
inter-Landau-band mixing} According to the semiclassical
theory\cite{chalker}, the motion of an electron in a strong
magnetic field and in a smooth random potential can be decomposed
into a rapid cyclotron motion and a slow drifting motion of its
guiding center. The kinetic energy of the cyclotron motion is
quantized by $E_n=(n+1/2)\hbar\omega_c$, where $\omega_c$ is the
cyclotron frequency and $n$ the LB index. The trajectory of the
drifting motion of the guiding center is thus along an
equipotential contour of value $V_0=E-E_n$, where $E$ is the total
energy of the electron. The local drifting velocity
$\vec{v}(\vec{r})$ is determined by (in SI unit)
\begin{equation}
\vec{v}(\vec{r})=\bigtriangledown V
(\vec{r})\times\vec{B}/(eB^2)
\label{drift}
\end{equation}
where $\bigtriangledown V(\vec{r})$ is the local potential
gradient. The equipotential contour consists of many closed loops.
Neglect quantum tunneling effects, each loop corresponds to
trajectory of one eigenstate. The motion of electrons are thus
confined around these loops with deviations typically order of the
cyclotron radius $l_c=\sqrt{\hbar/(eB)}$.
\begin{figure}[ht]
\begin{center}
  \includegraphics[height=4.4cm, width=8cm]{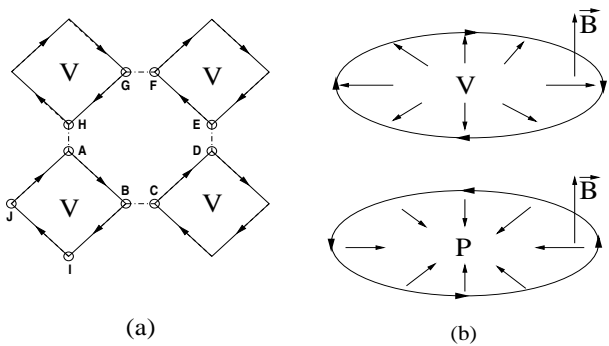}
\end{center}
\caption{(a) Four neighboring loops in a one-band model for the
case of $V_0<0$. Dashed lines denote quantum tunnelings. The
arrows indicate the drifting direction. (b) Loops localized around
a valley and a peak, respectively. The arrows inside a loop show
the directions of local potential gradient around the peak or
valley. The arrows on a loop indicate the drifting direction.}
\label{1channel}
\end{figure}

To illustrate this semiclassical picture, let us think of the
smooth random potential as a landscape of many peaks and valleys
distributed randomly in the plane. Imagine that the landscape is
filled with water up to a surface with the height of value $V_0$.
The equipotential contour of value $V_0$ is thus the intersection
between the land and the water. According to the percolation
theory\cite{stauff}, the percolation threshold of a
two-dimensional (2D) continuum model is $p_c=1/2$, where $p_c$ is
the occupation probability of the medium (the land or the water).
For simplicity, we suppose that the distribution of the random
potential is symmetric around zero. By symmetry the percolation
point of both the land and the water is at $V_0=0$ in this case.
When $V_0<0$, the occupation probability of land is above $1/2$.
Thus the land percolates and the water forms isolated lakes. These
lakes are around valleys and their boundaries correspond to
trajectories of localized states. In the case of $V_0>0$, the
water forms a percolating sea and the land becomes isolated
islands around potential peaks. The boundary of each island is an
electronic state. In short, semiclassical electronic states in a
QH system are equipotential loops. These loops are localized
around potential peaks for $V_0>0$ and around potential valleys
for $V_0<0$. The drifting direction of each loop is {\it
unidirectional}. This means that they are chiral states. From Eq.
\ref{drift} one can see that states around a peak have {\it
opposite chirality} from states around a valley because the
directions of the local potential gradients around a peak are
opposite to those around a valley. If one views the plane from the
direction opposite to the magnetic field, the drifting is
clockwise around valleys and counter-clockwise around peaks, as
shown in Fig. \ref{1channel}. Right at $V_0=0$ both the land and
the water percolate, and the intersection between them is the
trajectory of an extended state. It means that there is only one
extended state at $V_0=0$ for each LB. As $V_0$ approaches zero
from both sides, the localization length $\xi$ of the system tends
to diverge as
\begin{equation}
    \xi\propto |V_0|^{-\nu}
\label{ll}
\end{equation}
where the critical exponent $\nu=4/3$ according to the classical
percolation theory. Quantum effects are ignored in the above
semiclassical argument. When two spatially separated loops on the
same equipotential contour come close at saddle points of the
random potential, quantum tunnelings should be considered. An
example in the case of $V_0<0$ is shown in Fig. \ref{1channel}(a).
In the absence of interband mixing, numerical calculations have
suggested that there is still only one extended state in each LB
while the value of the critical exponent $\nu$ is modified to be
around $7/3$\cite{galstyan}.
\begin{figure}[ht]
\begin{center}
  \includegraphics[height=7cm,width=8cm]{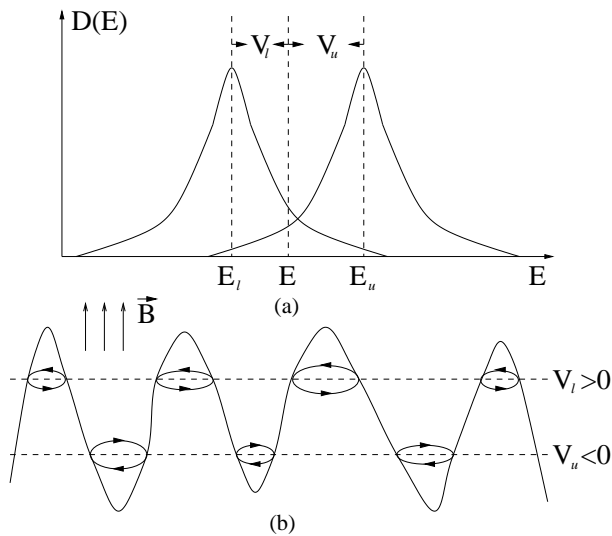}
\end{center}
\caption{(a) Two adjacent Landau bands in the case when the
disorder broadened band width is comparable with the Landau
gap. $D(E)$ is the density of states. $E_{\it u}$ and
$E_{\it l}$ denote the centers of the two bands.
(b) Two sets of equipotential contours for electronic
states of energy $E$ shown in (a). One is from the lower
band with $V_0= V_{\it l}>0$, and the other from the upper
band with $V_0=V_{\it u}<0$. The ellipses denote the loops
and arrows on them show drifting directions. The solid
curve is the schematic plot of the random potential.
Two dashed horizontal lines indicate two constant potential
planes. $\vec{B}$ is the magnetic field.}
\label{twoband}
\end{figure}

\begin{figure}[ht]
\begin{center}
 \includegraphics[height=4.4cm, width=8cm]{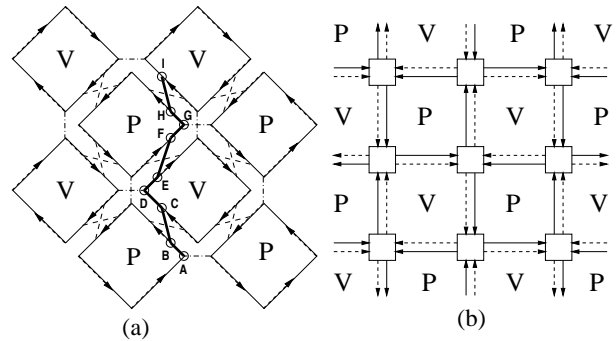}
\end{center}
\caption{(a) Topological plot of the trajectories of the drifting
motion of guiding centers (rhombus). The drifting motion around a
potential peak (valley) is denoted by P (V), and their direction
are indicated by the arrows. Dashed lines stand for interband
mixing, and dotted lines for tunneling at saddle points. The thick
line (A to I) describes the trajectory of an electron due to a
strong interband mixing. (b) The equivalent two-channel network
model of (a). Solid and dashed lines on each link denote two
channels from two LBs. Squares stand for saddle points. P, V and
arrows have the same meaning as those in (a).} \label{network}
\end{figure}

In the case of strong disorders or weak magnetic field, the width
of the LBs is comparable with the spacing between adjacent LBs
(the Landau gap), and inter-Landau-band mixings should no longer
be ignored. In order to investigate the consequences of
inter-Landau-band mixing, we shall consider a simple system of two
adjacent LBs. Since we are interested in interband-mixing of
opposite chirality, we consider those states with energy between
lower and upper bands which are centered at $E_l$ and $E_u$,
respectively, as shown in Fig. \ref{twoband}(a). Thus,
equipotential loops are $V_l=E-E_l>0$ and $V_u=E-E_u<0$ for the
lower and upper LBs, respectively. Using the semiclassical theory
described in the previous paragraphs, states from the upper band
should move along equipotential loops around potential valleys
while those from the lower band around potential peaks, as shown
in Fig. \ref{twoband}(b). The loops for the upper band drift in
clockwise direction, and those for the lower band in the
counter-clockwise direction. These two sets of loops are thus
spatially separated and have {\it opposite chirality}. If we
assume that the peaks and valleys of random potential form two
coupled square lattices, the loops can be arranged as shown in
Fig. \ref{network}(a), where {\it P} and {\it V} denote peaks and
valleys, respectively. In the absence of interband mixing, the
model is reduced to two decoupled single-band models and all
electronic states between the two LBs are localized. If we
introduce interband mixing, the localized loops may become less
localized. To see that this indeed occurs, let us consider an
extreme case with no tunnelings at saddle points, but with such
strong interband mixing that an electron will move from a loop
around a valley to its neighboring loop around a peak and vice
versa, as shown by $B\to C$ in Fig. \ref{network}(a). Follow an
electron starting at A, its trajectory will be $A\to B\to C\to
D\to E \cdot \cdot \cdot$. The electron is no longer confined on a
closed loop, but is now delocalized!

In the one-band model, an electron can also hop from one loop to
its neighboring loops by quantum tunnelings. At a first glance,
this effect seems similar to that of interband-mixing. However,
they are fundamentally different. In the one-band model,
electronic states for a given $V_0$ are of {\it the same
chirality}. Thus the drifting direction of an electron will be
inverted when it tunnels into neighboring loops. This means that
strong tunnelings in a one-band model will induce an effective
backward-scattering which also localizes the electrons. We can
understand this by considering a small part of the one-band model
as shown in Fig. \ref{1channel}(a) where all loops are moving in
clockwise direction. Without tunneling, the trajectory of an
electron starting from point A is $A\to B\to I\to J\to A$, a
clockwise closed loop. With strong tunnelings, the trajectory will
tend to be $A\to B\to C\to D \cdot \cdot \cdot \to H\to A$, a
counter-clockwise closed loop. Thus, the tunnelings between loops
of the same chirality cannot delocalize states.

It is worthwhile to explain why we consider only those states
between two LB centers. For states outside this range, both sets
of loops are localized around either valleys or peaks. This means
that interband mixing mainly occurs between two loops localized
around the same position, and this mixing will not delocalize a
state. In fact, as explained in the previous paragraph, the mixing
of the same chirality does not help delocalize an electron. This
is why we shall consider mixing between spatially separated states
with opposite chirality. Of course, it does not mean that the
mixing of the same chirality has no effect at all. As it was found
in some previous works\cite{wang}, this kind of mixing may shift
an extended state from its LB center. Level shifting due to mixing
between states of the same chirality may distort the shape of the
phase diagram, but should not alter its topology. The emergence of
the bands of extended states is exclusively due to the mixing
between states of opposite chirality.

\begin{figure}[ht]
\begin{center}
  \includegraphics[height=5.0cm,width=8cm]{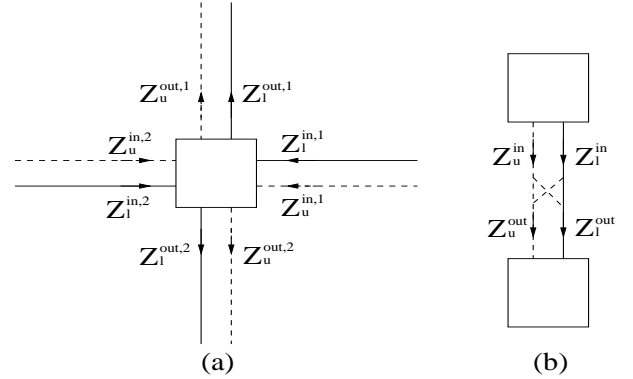}
\end{center}
\caption{(a) A node with four incoming and four outgoing channels.
$Z_{u(l)}^{in,i}$ is the wavefunction amplitude of the i$^{th}$
incoming wave from the upper (lower) LB. $Z_{u(l)}^{out,i}$ is
that of outgoing wavefunction amplitude. (b) A link with two
channels. $Z_{u(l)}^{in (out)}$ is the incoming (outgoing)
wavefunction amplitude of the upper (lower) LB.} \label{nodelink}
\end{figure}

Now, we describe our two-channel network model in detail. Assume
that tunnelings of two neighboring localized states (loops) of the
same band occur around saddle points, and interband mixing takes
place only on the links, Fig. \ref{network}(a) is topologically
equivalent to the model shown in Fig. \ref{network}(b). Fig.
\ref{network}(b) is the schematic illustration of our {\it
two-channel Chalker-Coddington} network model. It is similar to
the model studied in previous publications\cite{lee, kagalovsky}.
There are two channels on each link. One, denoted by a solid line,
is from the lower LB around a potential peak. The other (dashed
line) is from the upper LB moving around a potential valley. The
arrows indicate the drifting direction of the two sets of states.
At each node, the tunneling between two neighboring states of the
same LB occurs. As shown in Fig. \ref{nodelink}(a), let
$Z_{u(l)}^{in,1}$ and $Z_{u(l)}^{in,2}$ be the incoming wave
amplitudes of states 1 and 2 from upper (lower) LB, respectively,
and $Z_{u(l)}^{out,1}$ and $Z_{u(l)}^{out,2}$ be the outgoing wave
amplitudes of the two states. The tunneling is described by a
SO(4) matrix.
\begin{equation}
\label{smatrix}
    \left (
    \begin{array}{l}
    Z_u^{out,1}\\
    Z_u^{out,2}\\
    Z_l^{out,1}\\
    Z_l^{out,2}
    \end{array}
    \right )
    =
    \left (
    \begin{array}{llll}
    s_u^R & s_u^L & 0 & 0 \\
    -s_u^L & s_u^R & 0 & 0 \\
    0 & 0 & s_l^R & s_l^L \\
    0 & 0 & -s_l^L & s_l^R
    \end{array}
    \right )
    \left (
    \begin{array}{l}
    Z_u^{in,1}\\Z_u^{in,2}\\Z_l^{in,1}\\Z_l^{in,2}
    \end{array}
    \right ) ,
\end{equation}
where the subscripts $u$ and $l$ denote the upper and lower bands,
respectively. The elements $s_{u(l)}^L$ and $s_{u(l)}^R$ are
tunneling coefficients of an incoming wave-function in the upper
(lower) band being scattered into outgoing channels at its
left-hand and right-hand sides, respectively. $s_{u(l)}^R$ and
$s_{u(l)}^L$ are related to each other as $s_{u(l)}^R=\sqrt
{1-(s_{u(l)}^L)^2}$ due to the orthogonality of the matrix. Under
quadratic potential barrier approximation, ---i.e.,
$V(x,y)=-Ux^2+Uy^2+V_c$ around a saddle point, where $U$ is a
constant describing the strength of potential fluctuation and
$V_c$ is the potential barrier at the point,--- one can show that
the left-hand scattering amplitude is given by\cite{fertig1}
\begin{equation}
    s_{u(l)}^L=[1+\exp(-\pi\epsilon_{u(l)})]^{-1/2},
\end{equation}
where $\epsilon_{u(l)}=[E+V_c-(n_{u(l)}+1/2)E_2]/E_1$ with $E$ the
electronic energy, $E_1=\frac{\hbar\omega_c}{2\sqrt{2}}\sqrt{K-1}$
and $E_2=\frac{\hbar\omega_c}{\sqrt{2}}\sqrt{K+1}$ with
$K=\sqrt{\frac{64U^2}{m^2\omega_c^4}+1}$. The kinetic energies of
cyclotron motion in the two bands are $(n_u+1/2)E_2$ and
$(n_l+1/2)E_2$, respectively, where $n_{u(l)}$ are the band
indices and $\Delta n=n_u-n_l=1$. The dimensionless ratio
$E_r=E_2/E_1=2\sqrt {1+\frac{2}{K-1}}$ approaches $2$ from above
as $U$ or the inverse of $\omega_c$ increases\cite{fertig1}, i.e.,
the regime of strong disorders or weak magnetic field. Since this
is the regime we are interested in, we choose the value of it to
be 2.2 in our calculations. For convenience, we choose $E_2$ as
the energy unit and the cyclotron energy of the lower band as the
reference point. The energy regime between the two band centers is
thus $E\in[0,1]$.

Inter-band mixing between two channels on a link as shown
in Fig. \ref{nodelink}(b) is described by a U(2) matrix
\begin{equation}
\label{mmatrix1}
    \left (
    \begin{array}{l} Z^{out}_l\\Z^{out}_u \end{array}
    \right )=M
    \left (
    \begin{array}{l} Z^{in}_l\\Z^{in}_u \end{array}
    \right ),
\end{equation}
\begin{equation}
\label{mmatrix2}
    M=\left (
    \begin{array}{ll} e^{i\phi_1} & 0 \\ 0 & e^{i\phi_2}
    \end{array}
    \right )
    \left (
    \begin{array}{ll} \cos\theta & \sin\theta \\ -
    \sin\theta & \cos\theta \end{array}
    \right )
    \left (
    \begin{array}{ll} e^{i\phi_3} & 0 \\ 0 & e^{i\phi_4}
    \end{array}
    \right ),
\end{equation}
where $\sin\theta$ describes the interband mixing. $\phi_i(i=1\sim
4)$ are random Aharonov-Bohm phases accumulated along propagation
paths. In our calculations, we shall assume that they are
uniformly distributed in $[0,2\pi]$\cite{chalker}. In the
following discussion, a parameter $P$, defined as
$\sqrt{P/(1+P)}=\sin\theta$, is used to characterize the mixing
strength. $P$ will take the same value for all links in our
calculations. We hope that this simplification will not affect the
physics.

\section{The application of level-statistics technique
on the network model}

Electron localization length is often obtained from the transfer
matrix method. For a two-dimensional system, however, it is well
known that this quantity alone does not provide conclusive answers
to questions related to the metal-insulator transition
(MIT)\cite{xie}. On the other hand, level-statistics analysis has
been used in studying MIT\cite{kramer,schweitzer}.
Level-statistics analysis is based on random matrix theory
(RMT)\cite{mehta}. The basic idea is that the localization
property of an electronic state can be determined by the
statistical distribution function $P(s)$ of the spacing $s$ of two
neighboring levels. For localized states, the distribution is
Poissonian $P_{PE}(s)=exp(-s)$, called `Poissonian ensemble (PE)'.
In the case of extended states, the nearest neighbor level spacing
distribution has the following form\cite{mehta}
\begin{equation}
 P(s)=C_1(\beta)s^{\beta}exp[-C_2(\beta)s^2]
\end{equation}
where $C_1(\beta)$ and $C_2(\beta)$ are normalization factors
determined by $\int P(s)ds=1$ and $\int sP(s)ds=1$. The parameter
$\beta$ is determined by the dynamical symmetry of the system. The
case of $\beta=1$ is for systems with time-reversal symmetry and
an integer total angular momentum and is referred as `Gaussian
orthogonal ensemble'. Systems with time-reversal symmetry and a
half-integer total angular momentum belong to the case of
$\beta=4$, called `Gaussian symplectic ensemble'. For systems
without time-reversal symmetry $\beta=2$, and it is called
`Gaussian unitary ensemble (GUE)'.

We shall follow the approach proposed by Klesse and
Metzler\cite{klesse}. A quantum state of a network model can be
expressed by a vector whose components are electronic
wave-function amplitudes on the links. In our case, the vector can
be written as $\Phi=(\{Z_u^i,Z_l^i\})$, where $Z_u^i$ and $Z_l^i$
are the electron wave-function amplitudes of the upper band ({\it
u}) and the lower band ({\it l}) on the $i$-th link, respectively.
As shown by Fertig\cite{fertig2}, the network model can be
described by an {\it evolution operator} $\hat{U}(E)$, an
$E$-dependent matrix determined by the scattering properties of
nodes and links in the model. (As an example, the evolution
operator of a two-channel network of size $L=2$ with periodic
boundaries on both directions is constructed explicitly in the
Appendix.) In general, the eigenvalue equation of the evolution
operator is
\begin{equation}
\hat{U}(E)\Phi_{\alpha}(E)=
e^{i\omega_{\alpha}(E)}\Phi_{\alpha}(E),
\end{equation}
where $\alpha$ is the eigenstate index of $\hat U$. The true
eigenenergies $\{E_n\}$ of the system are those energies at which
$\omega_\alpha(E)$ is an integer multiple of $2\pi$. It has been
shown by Klesse and Metzler\cite{klesse} that the level-spacing
statistics of the set of {\it quasi-energies}
$\{\omega_{\alpha}(E_{n})\}$ is the same as that of $\{E_n\}$.
Thus the localization property of an electronic state with an
energy $E$ can be obtained by the quasi-energies. The advantage of
this approach is that all the quasi-energies can be used in the
analysis so that better statistics can be obtained.

Chalker and Coddington\cite{chalker} showed numerically that an
open boundary condition along one direction creates extended edge
states along the other direction. In order to get rid of the edge
states, we employ a periodic boundary along both directions in our
calculation. For a two-channel network model of $L\times L$ nodes
with periodic boundaries along both directions, there are $4L^2$
components in $\Phi$. $\hat{U}$ is thus a $(4L^2)\times (4L^{2})$
matrix. However, there is a special property of the network
model\cite{metzler}: the nodes scatter electrons only from
vertical channels into horizontal channels and {\it vice versa}.
If one separates $\Phi$ into the set of wavefunction amplitudes on
the horizontal links $\Phi_H$ and the set of wavefunction
amplitudes on the vertical links $\Phi_V$, the evolution equation
in one time step can be written in the following form
\begin{equation}
\label{evolution}
    \left (
    \begin{array}{l}
    \Phi_H(t+1) \\ \Phi_V(t+1)
    \end{array}
    \right )
    =
    \left (
    \begin{array}{ll}
    \hat{0} & \hat{U}_{V\to H} \\
        \hat{U}_{H\to V} & \hat{0}
    \end{array}
    \right )
    \left (
    \begin{array}{l}
    \Phi_H(t) \\ \Phi_V(t)
    \end{array}
    \right ),
\end{equation}
where $\hat{0}$ is the $(2L^2)\times (2L^{2})$ zero matrix.
$\hat{U}_{V\to H}$ describes how wavefunction on vertical links
evolves into that on the horizontal  links. Similarly,
$\hat{U}_{H\to V}$ describes that from horizontal to vertical
links. For the detail derivation, we refer readers to the example
shown in the Appendix. The evolution equation in two time steps is
given as
\begin{eqnarray}
\Phi_H(t+2)=
\hat{U}_{V\to H}\hat{U}_{H\to V}\Phi_H(t) \\
\Phi_V(t+2)=
\hat{U}_{H\to V}\hat{U}_{V\to H}\Phi_V(t)
\end{eqnarray}
Therefore, the evolution matrix in two time steps is
block-diagonal and the two blocks have essentially the same
statistical property. We thus need only consider one of them.

We study the model for L=8, 12, 16,  20 and 24. The calculation
procedure is as follows. Take a realization of the random phases,
construct the evolution matrix and obtain the quasi-energies
$\{\omega_{i}\}$. Put them in descending order and calculate the
level spacings $s_i=(\omega_i-\omega_{i-1}) /\delta$, where
$\delta$ is the average of ${s_i}$. Repeat this procedure for
sufficient times so that more than $5\times10^4$ level spacings
are collected for a given $E$ and $P$. The level-spacing
distribution function $P(s)$ is thus obtained numerically.

\section{Numerical results and discussions}
\subsection{Analysis of the level-spacing spectrum}

\begin{figure}[ht]
\begin{center}
  \includegraphics[width=8cm,height=4.9cm]{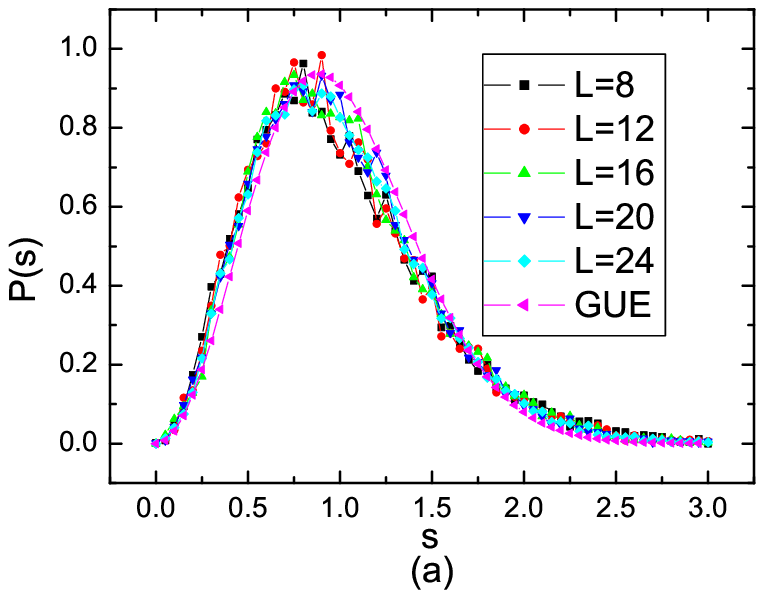}
\end{center}
\begin{center}
  \includegraphics[width=8cm,height=4.9cm]{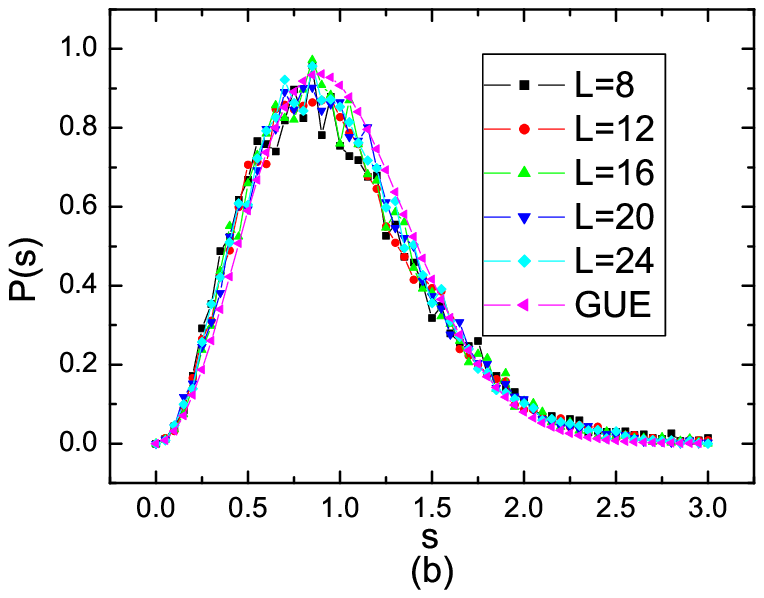}
\end{center}
\begin{center}
  \includegraphics[width=8cm,height=4.9cm]{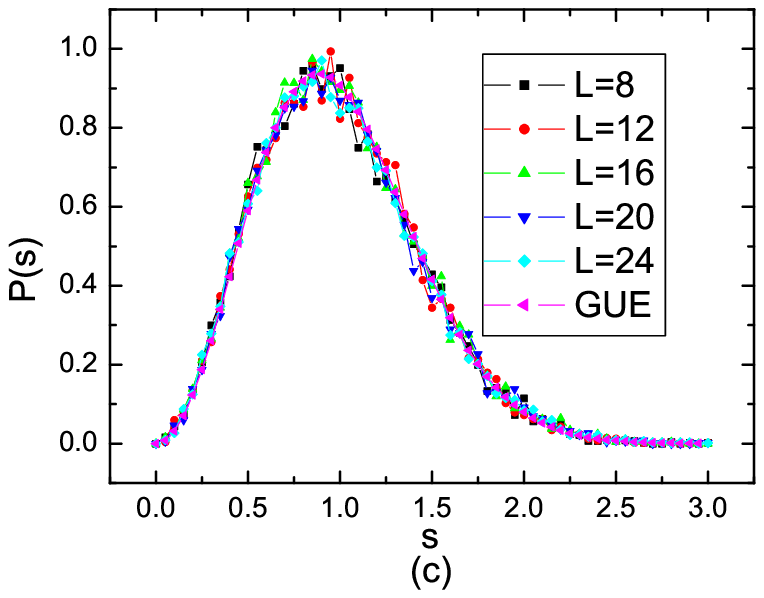}
\end{center}
\caption{$P(s)$ vs. $s$ for $L=8,12,16,20,24$ and comparison with
the Gaussian unitary ensemble $P_{GUE}(s)$. (a) $E=0$ and $P=0.1$;
(b) $E=0.02$ and $P=0.1$; (c) $E=0.5$ and $P=1.5$.}
\label{ps_global1}
\end{figure}

\begin{figure}[ht]
\begin{center}
  \includegraphics[width=8cm,height=4.9cm]{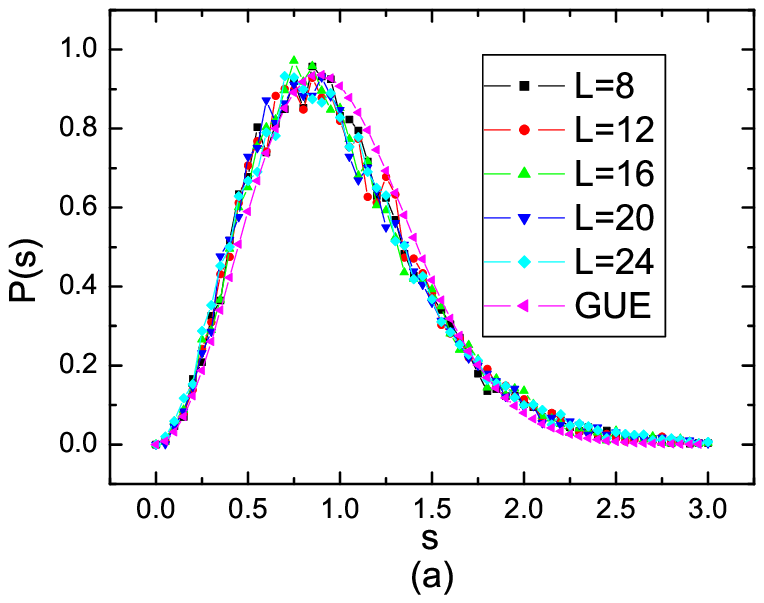}
\end{center}
\begin{center}
  \includegraphics[width=8cm,height=4.9cm]{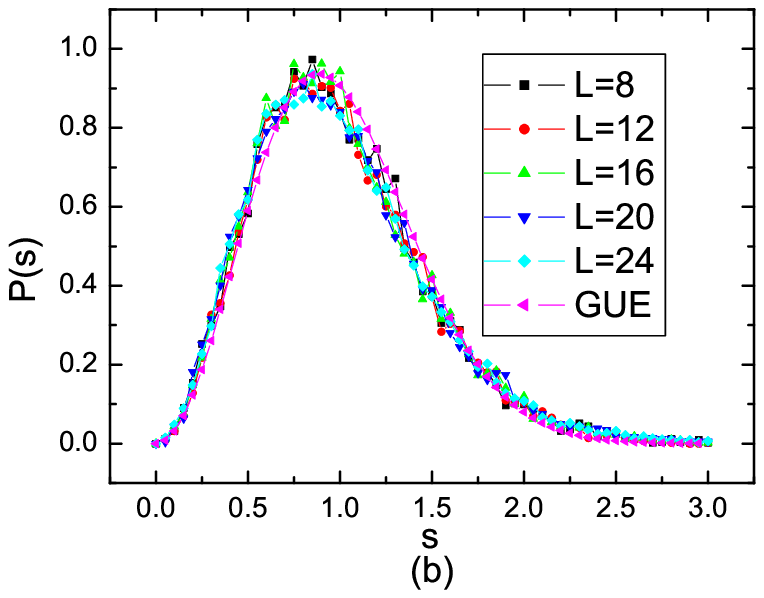}
\end{center}
\begin{center}
  \includegraphics[width=8cm,height=4.9cm]{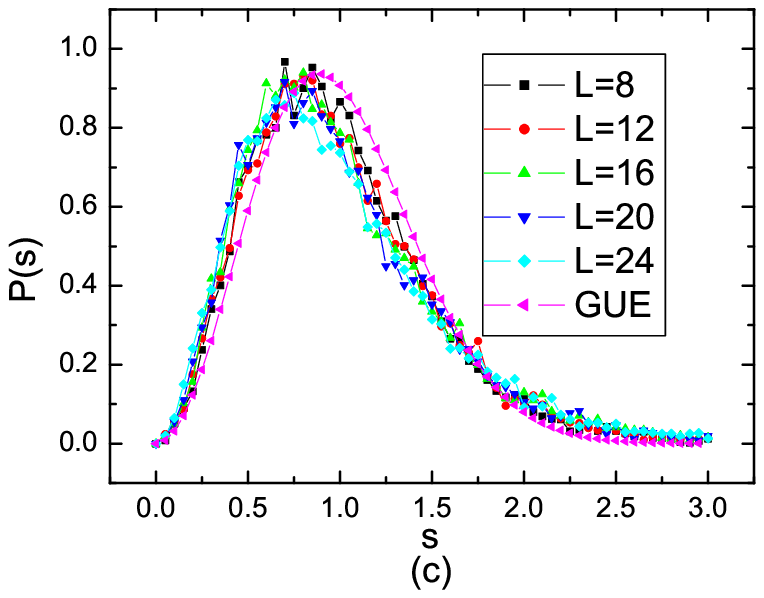}
\end{center}
\caption{$P(s)$ vs. $s$ for $L=8,12,16,20,24$ and comparison with
the Gaussian unitary ensemble $P_{GUE}(s)$. (a) $E=0$ and $P=0.7$;
(b) $E=0.02$ and $P=0.7$; (c) $E=0.5$ and $P=0.5$.}
\label{ps_global2}
\end{figure}

In the following, we shall analyze the numerical results of the
level-spacing distribution function $P(s)$. Our purpose is to show
evidence for the existence of bands of extended states in our
model. Due to the chiral nature of the drifting motion,
time-reversal symmetry is absent from our semiclassical network
model. Then, according to the RMT\cite{mehta}, if bands of
extended states exist, $P(s)$ of them should be the GUE
distribution $P_{GUE}(s)=32\pi^{-2}s^2exp(-4s^2/\pi)$. While
$P(s)$ of localized states is the PE distribution $P_{PE}(s)$.
Since the global shapes of GUE and PE are quite different, let us
first take a look at the global shape of our numerical results of
$P(s)$. Curves in Fig.\ref{ps_global1} show $P(s)$ at
$(E=0,P=0.1)$ (a), $(E=0.02,P=0.1)$ (b), and $(E=0.5,P=1.5)$ (c)
for $L=8,12,16,20,24$ and comparison with the Gaussian unitary
ensemble distribution $P_{GUE}(s)$, while Fig.\ref{ps_global2} is
for $(E=0.0,P=0.7)$ (a), $(E=0.02,P=0.7)$ (b) and $(E=0.5,P=0.5)$
(c).
\begin{figure}[ht]
\begin{center}
  \includegraphics[width=8cm,height=4.9cm]{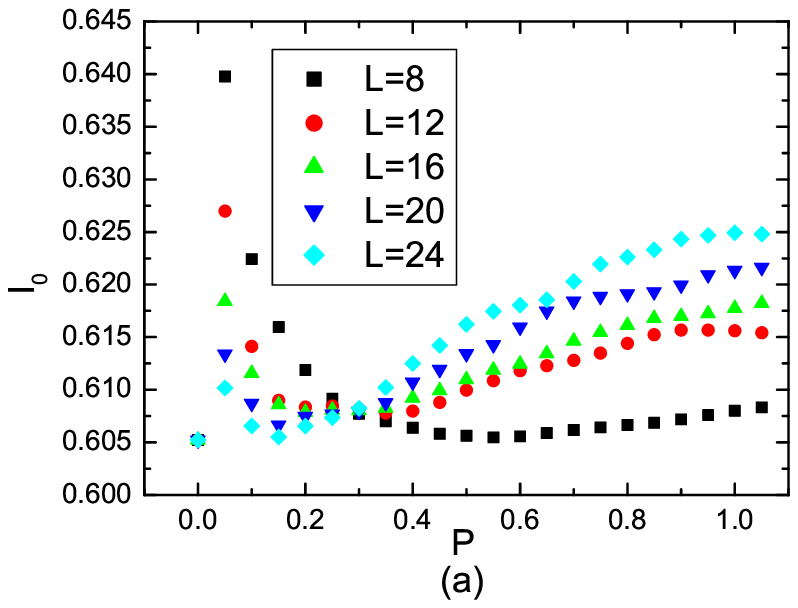}
\end{center}
\begin{center}
  \includegraphics[width=8cm,height=4.9cm]{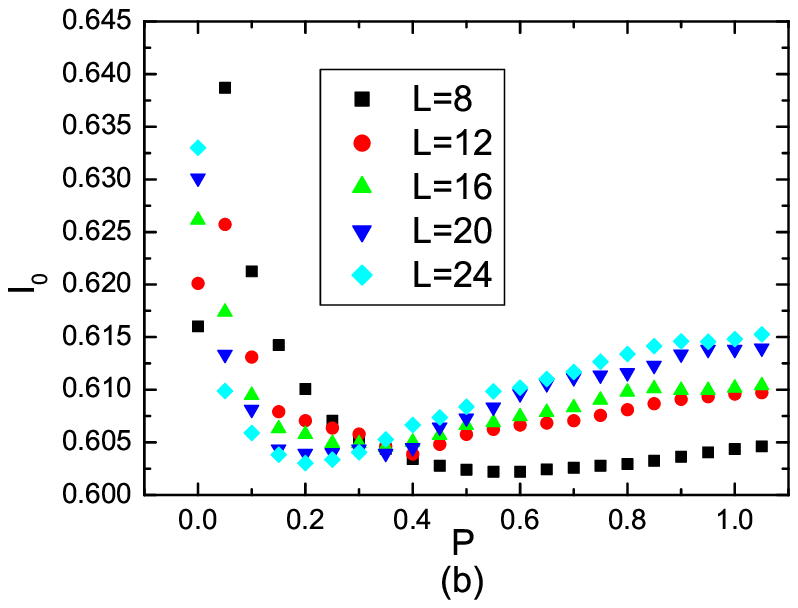}
\end{center}
\begin{center}
  \includegraphics[width=8cm,height=4.9cm]{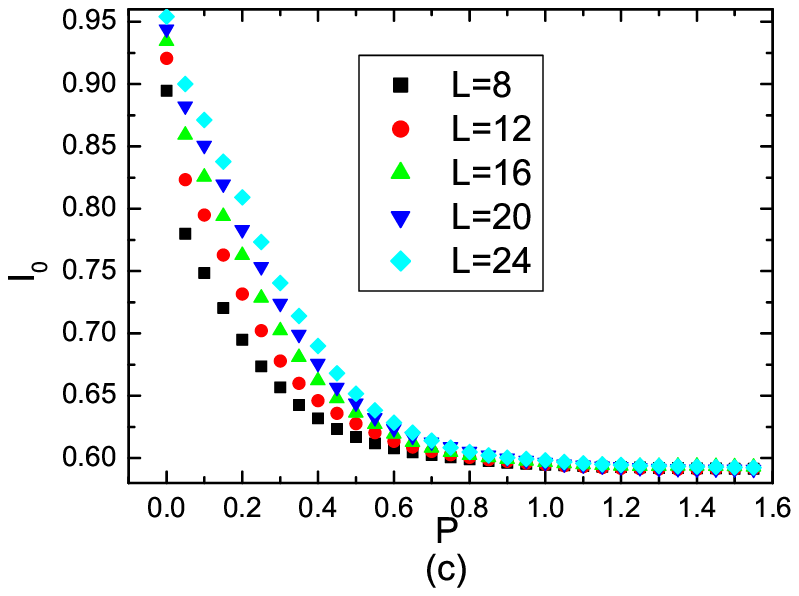}
\end{center}
\caption{$I_0$ vs. $P$ for $L=8, 12, 16, 20, 24$, (a) $E=0$; (b)
$E=0.02$; (c) $E=0.5$.} \label{data}
\end{figure}
\begin{figure}[ht]
\begin{center}
  \includegraphics[width=8cm,height=4.9cm]{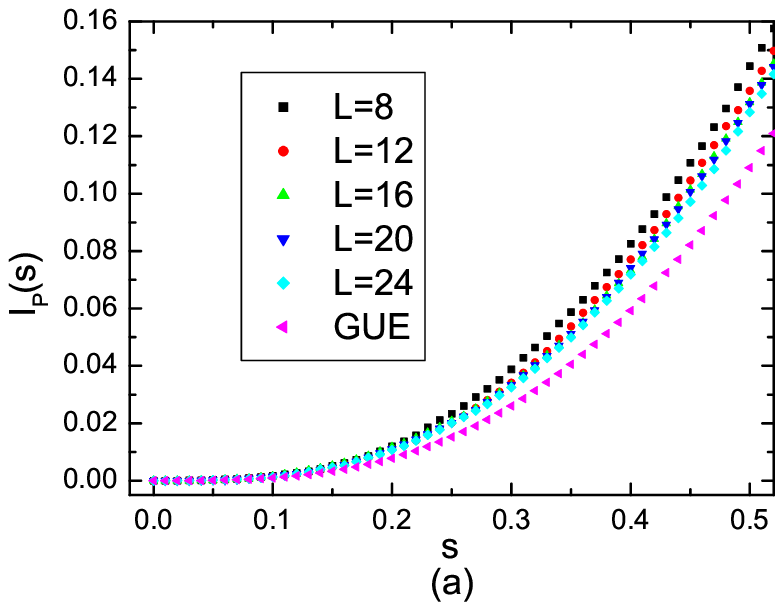}
\end{center}
\begin{center}
  \includegraphics[width=8cm,height=4.9cm]{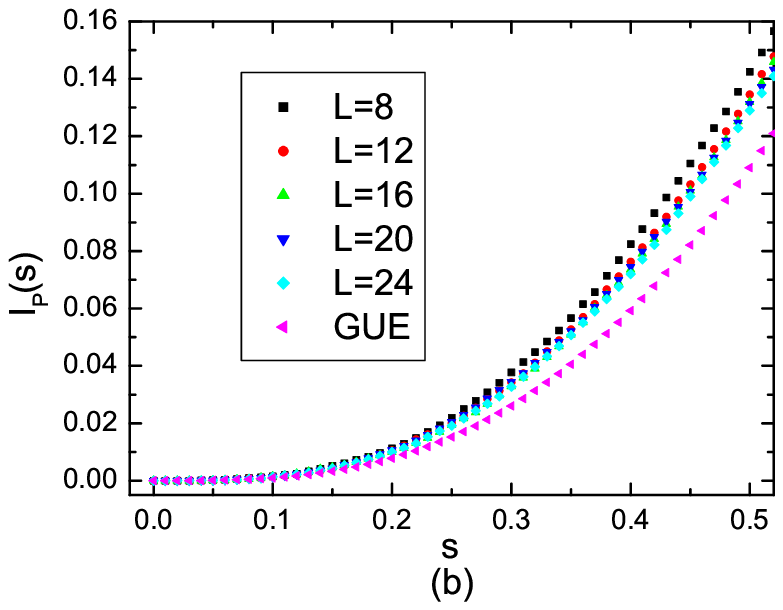}
\end{center}
\begin{center}
  \includegraphics[width=8cm,height=4.9cm]{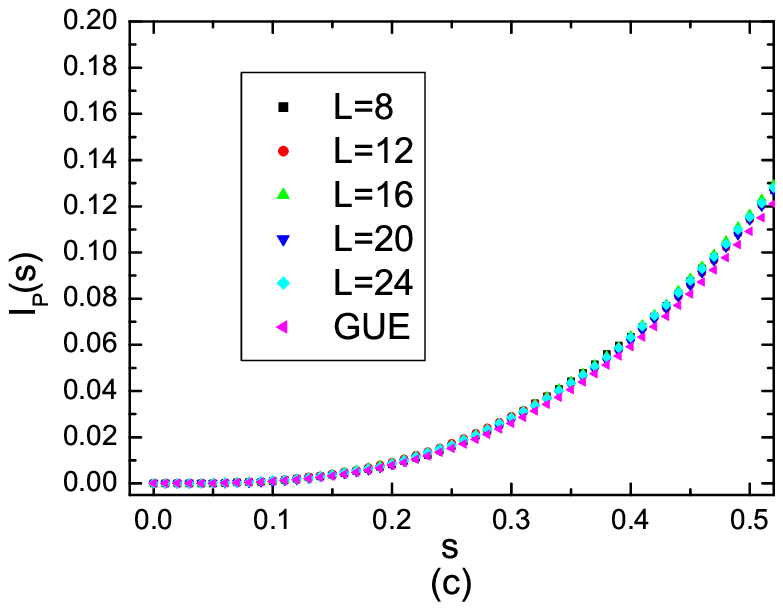}
\end{center}
\caption{$I_P(s)$ vs. $s$ for $L=8, 12, 16, 20, 24$ and comparison
with that of $P_{GUE}(s)$. (a) $E=0$ and $P=0.1$; (b) $E=0.02$ and
$P=0.1$; (c) $E=0.5$ and $P=1.5$.} \label{ps_small1}
\end{figure}
\begin{figure}[ht]
\begin{center}
  \includegraphics[width=8cm,height=4.9cm]{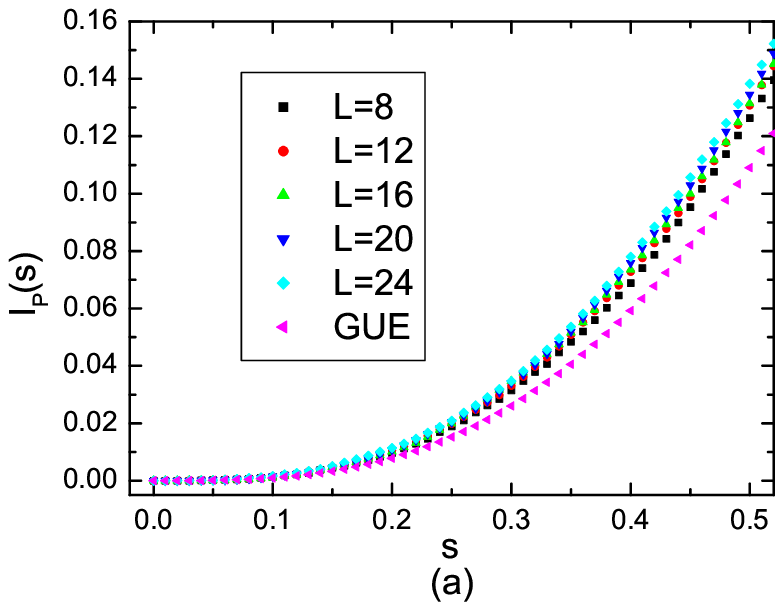}
\end{center}
\begin{center}
  \includegraphics[width=8cm,height=4.9cm]{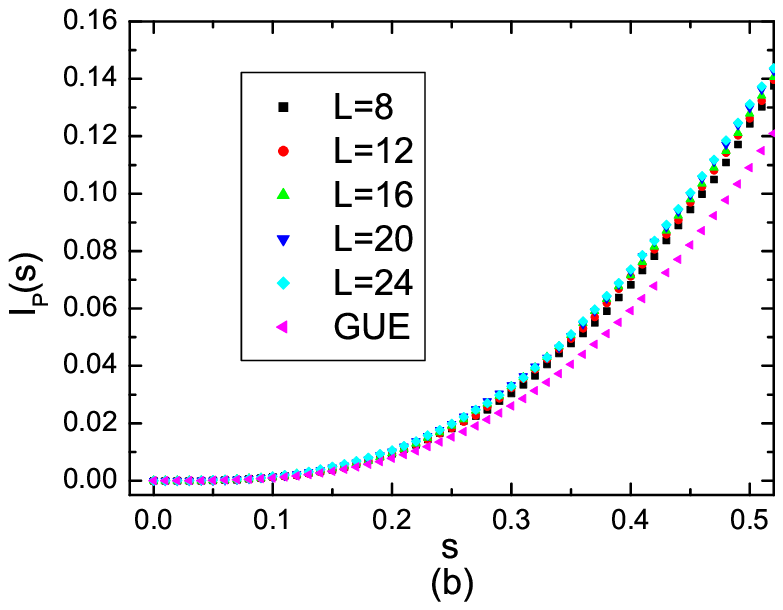}
\end{center}
\begin{center}
  \includegraphics[width=8cm,height=4.9cm]{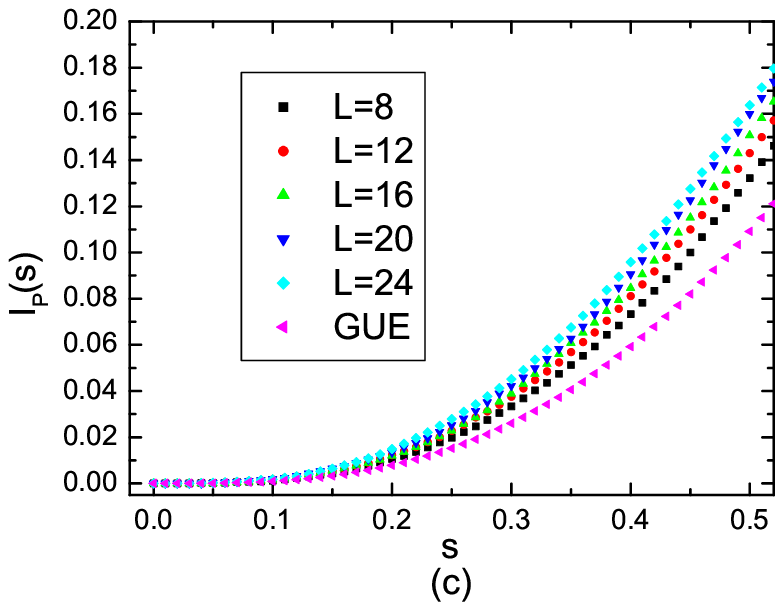}
\end{center}
\caption{$I_P(s)$ vs. $s$ for $L=8, 12, 16, 20, 24$ and comparison
with that of $P_{GUE}(s)$, (a) $E=0$ and $P=0.7$; (b) $E=0.02$ and
$P=0.7$; (c) $E=0.5$ and $P=0.5$.} \label{ps_small2}
\end{figure}
\begin{figure}[ht]
\begin{center}
  \includegraphics[width=8cm,height=4.9cm]{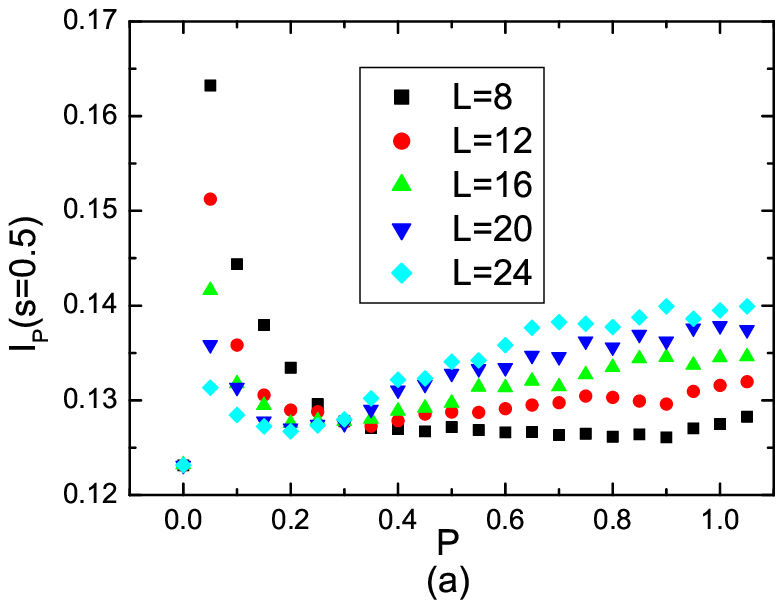}
\end{center}
\begin{center}
  \includegraphics[width=8cm,height=4.9cm]{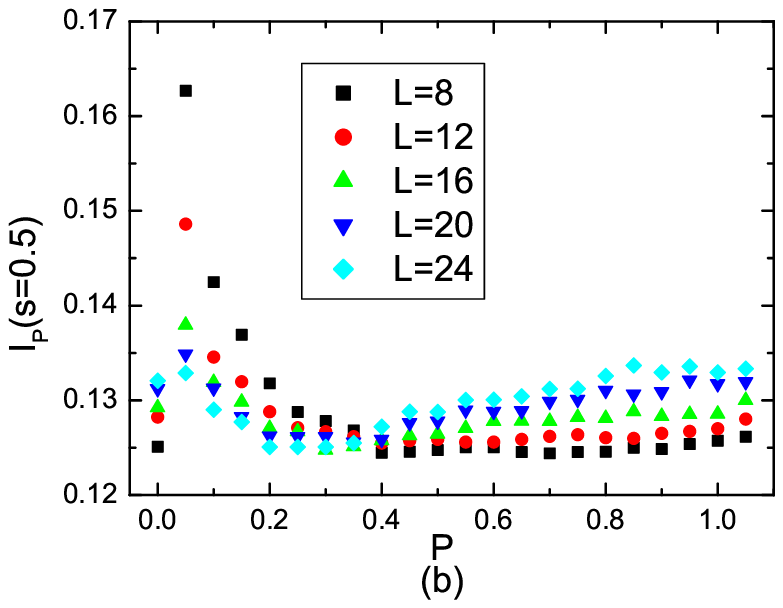}
\end{center}
\begin{center}
  \includegraphics[width=8cm,height=4.9cm]{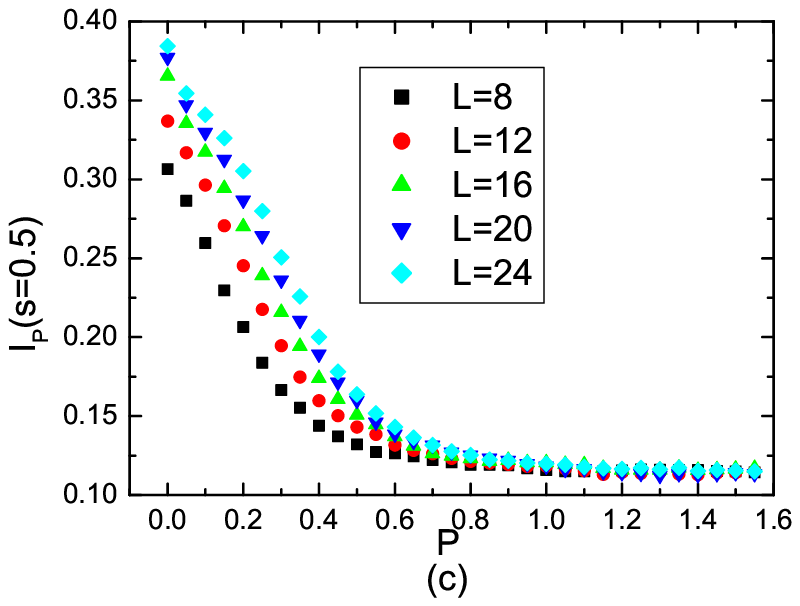}
\end{center}
\caption{$I_P(s=0.5,P)$ vs. $P$ for $L=8, 12, 16, 20, 24$, (a)
$E=0$; (b) $E=0.02$; (c) $E=0.5$.} \label{ps_small3}
\end{figure}
\begin{figure}[ht]
\begin{center}
  \includegraphics[width=8cm,height=4.9cm]{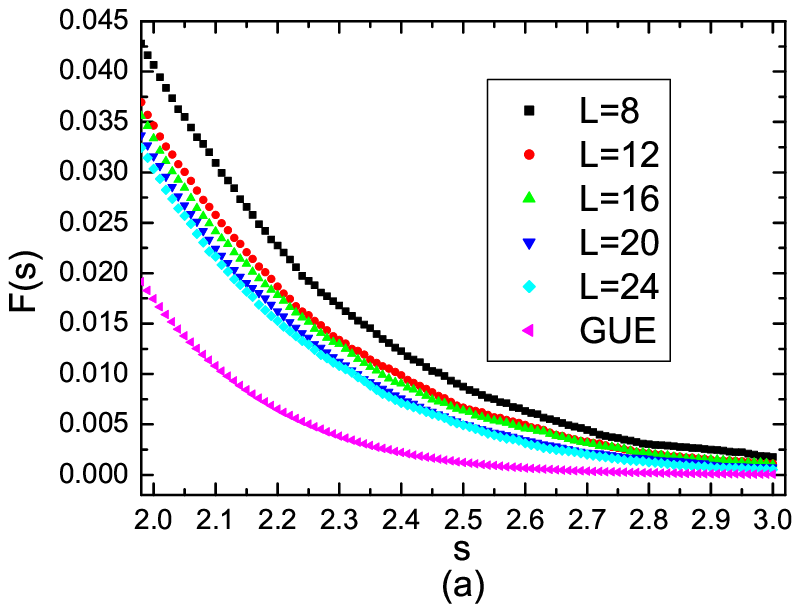}
\end{center}
\begin{center}
  \includegraphics[width=8cm,height=4.9cm]{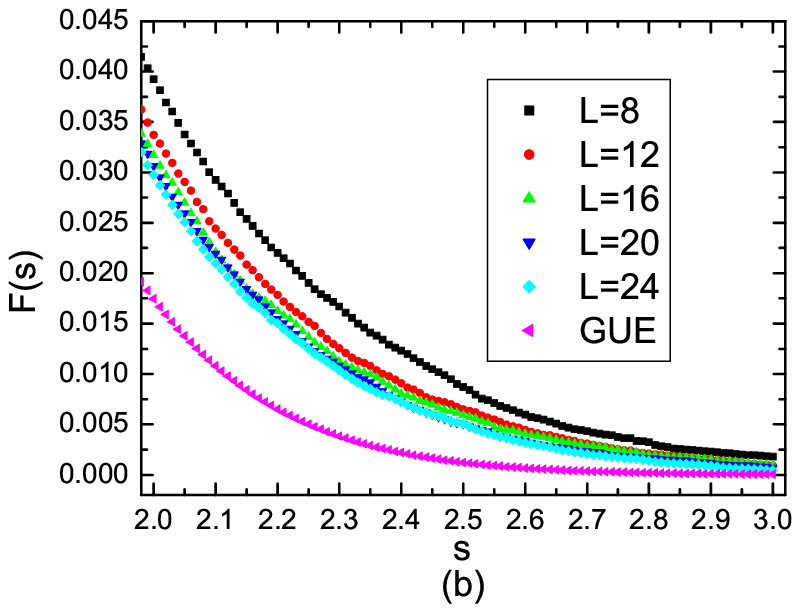}
\end{center}
\begin{center}
  \includegraphics[width=8cm,height=4.9cm]{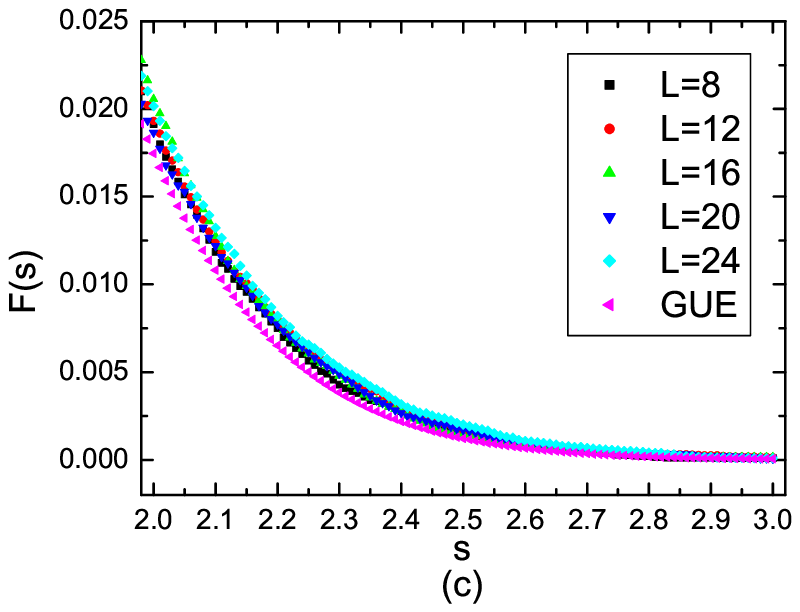}
\end{center}
\caption{$F(s)$ vs. $s$ for $L=8, 12, 16, 20, 24$, and comparison
with that of $P_{GUE}(s)$. (a) $E=0$ and $P=0.1$; (b) $E=0.02$ and
$P=0.1$; (c) $E=0.5$ and $P=1.5$.} \label{ps_tail1}
\end{figure}
\begin{figure}[ht]
\begin{center}
  \includegraphics[width=8cm,height=4.9cm]{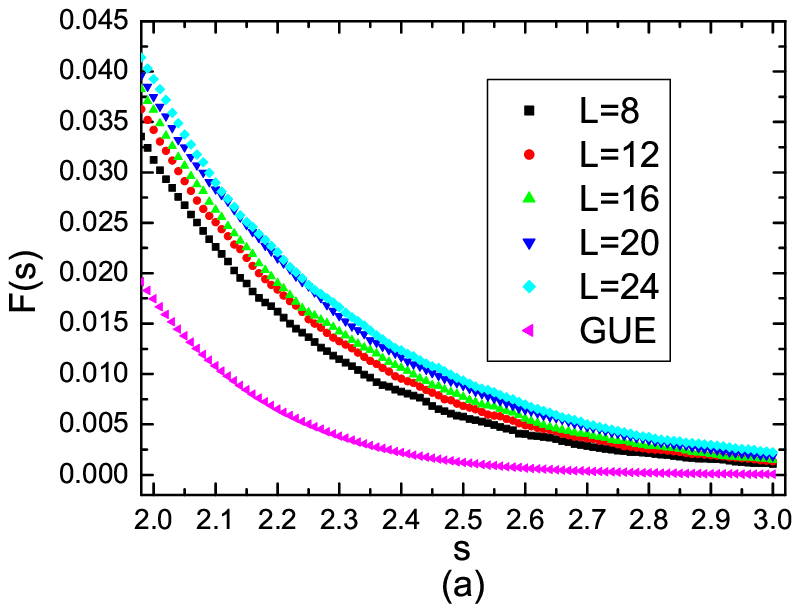}
\end{center}
\begin{center}
  \includegraphics[width=8cm,height=4.9cm]{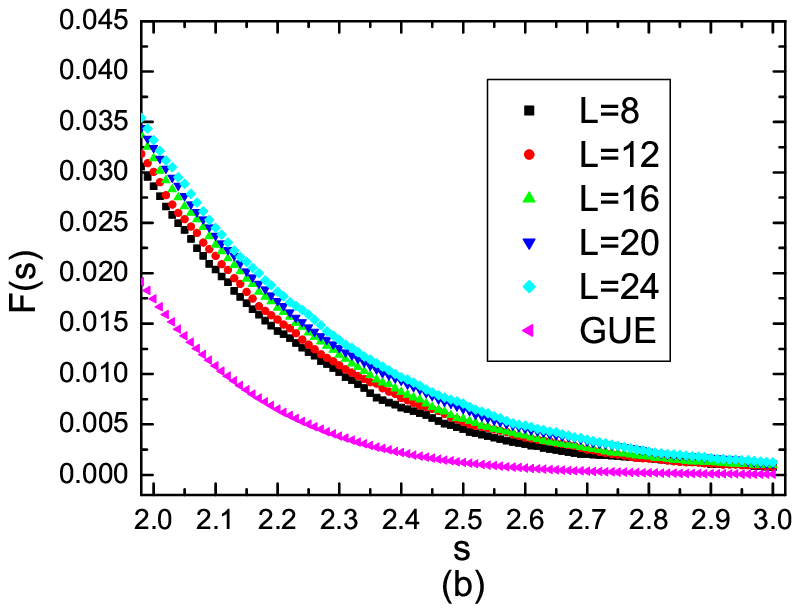}
\end{center}
\begin{center}
  \includegraphics[width=8cm,height=4.9cm]{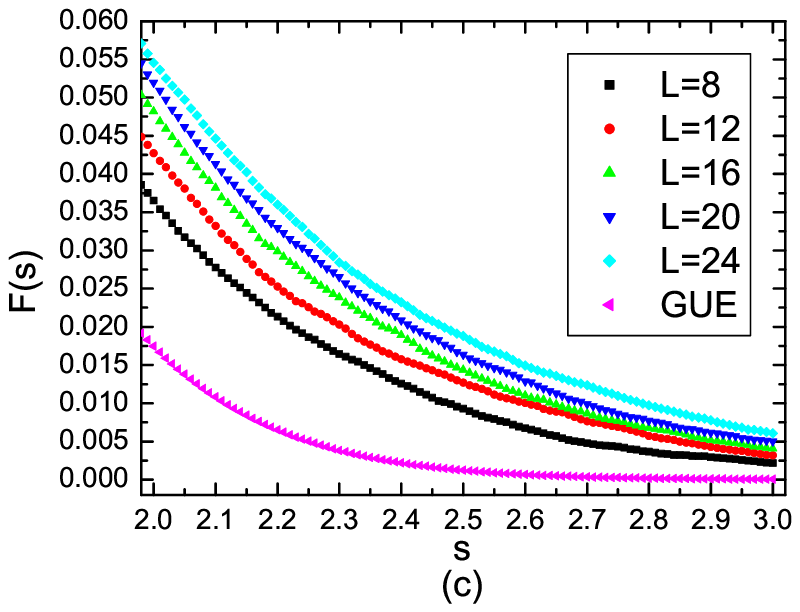}
\end{center}
\caption{$F(s)$ vs. $s$ for $L=8, 12, 16, 20, 24$ and comparison
with that of $P_{GUE}(s)$. (a) $E=0$ and $P=0.7$; (b) $E=0.02$ and
$P=0.7$; (c) $E=0.5$ and $P=0.5$.} \label{ps_tail2}
\end{figure}
\begin{figure}[ht]
\begin{center}
  \includegraphics[width=8cm,height=4.9cm]{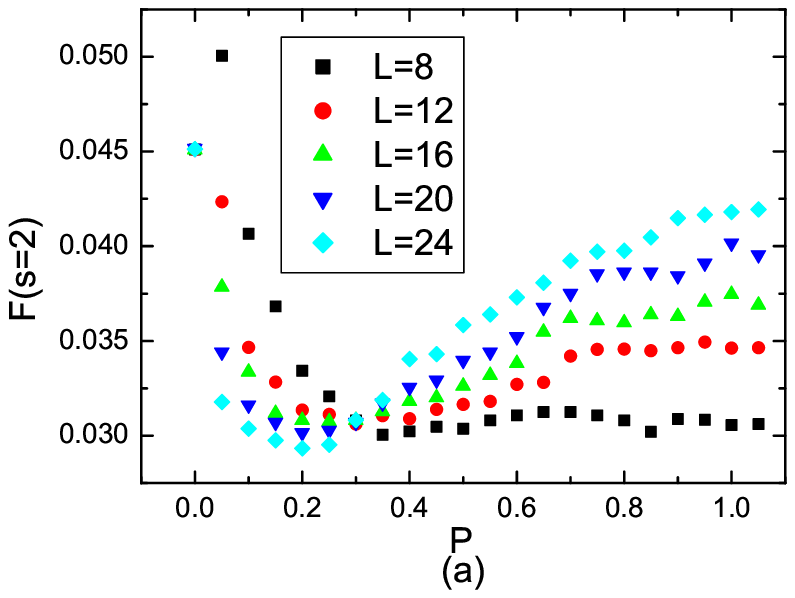}
\end{center}
\begin{center}
  \includegraphics[width=8cm,height=4.9cm]{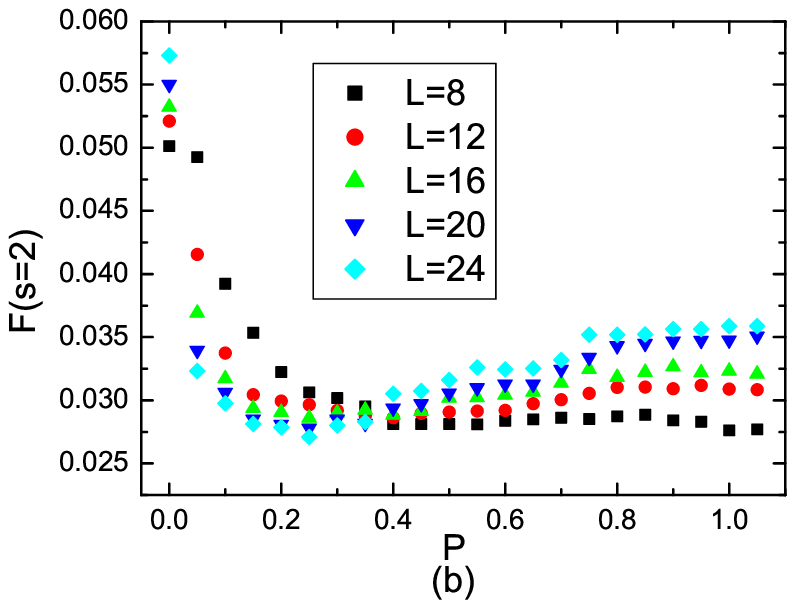}
\end{center}
\begin{center}
  \includegraphics[width=8cm,height=4.9cm]{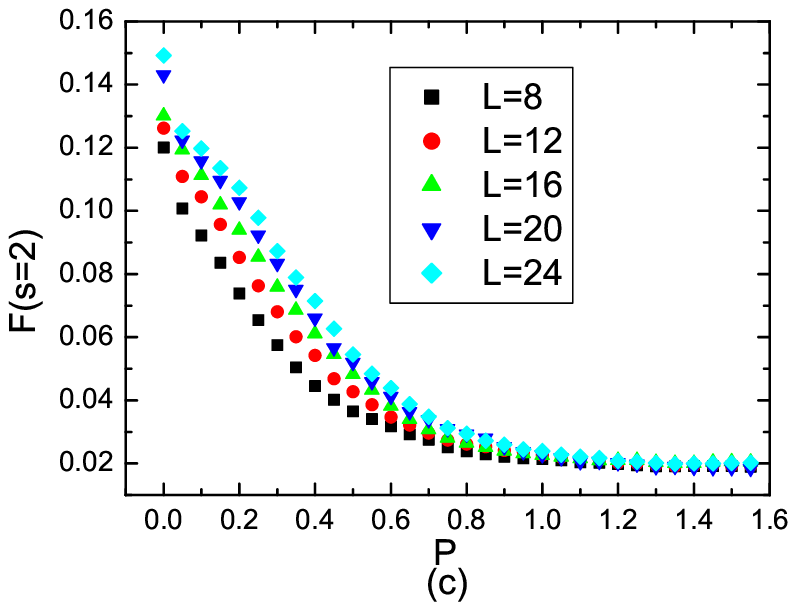}
\end{center}
\caption{$F(s=2,P)$ vs. $P$ for $L=8, 12, 16, 20, 24$, (a) $E=0$;
(b) $E=0.02$; (c) $E=0.5$.} \label{ps_tail3}
\end{figure}
The global shape of these curves has some common features. All
curves have a vanishing value when $s$ tends to zero. At small $s$
they increase with $s$ and reach a peak at some intermediate $s$.
Then they decrease with increasing $s$ and tend to vanish at large
$s$. All these features are the same as the GUE distribution
$P_{GUE}(s)$\cite{metzler}. Thus most of them are close to
$P_{GUE}(s)$ at first glance. This raises the question on how to
distinguish between extended states and localized states by our
numerical results. As a simple way, it is natural to expect that
$P(s)$ of extended states should tend closer to $P_{GUE}(s)$ or
remain unchanged with increasing $L$ while that of localized
states should deviate from $P_{GUE}(s)$ with increasing $L$. By
careful observation one can indeed see that curves in each
sub-figure of Fig.\ref{ps_global1} tend to be closer to
$P_{GUE}(s)$ with increasing $L$ while those in
Fig.\ref{ps_global2} show the opposite tendency. Thus we can use
the different {\it tendency} of $P(s)$ with increasing $L$ to
distinguish between extended states and localized states. In the
following, we shall show quantitatively such opposite tendencies
for extended states and localized states by considering several
characteristic quantities of $P(s)$.

Let us first consider a characteristic quantity $I_0$ defined by
$I_0=\int s^2P(s)ds/2$. It is commonly used to characterize the
global shape of $P(s)$ and to exam the localization
property\cite{metzler}. It is well-known that $I_0=1$ for
localized states while $I_0<1$ for extended states\cite{mehta}.
Thus, the following simple criteria is employed. If $I_0$ of a
state with energy $E$ increases and approaches $1$ with increasing
$L$, this state is localized. Otherwise, it is extended. Curves in
Fig.\ref{data} are $I_0$ vs. mixing strength $P$ at $E=0$ (a);
0.02 (b); and 0.5 (c) for $L=8, 12, 16, 20, 24$. Fig.\ref{data}(a)
shows that the state at $E=0.02$ is localized at zero mixing and
extended at small $P$. Then it is localized again after $P$ passes
a particular $P_c$ where $I_0$ of different $L$ cross. For the
state at the lower band center $E=0$ shown in Fig.\ref{data}(b),
it is extended at zero mixing. Then, it shows the same feature as
the state of $E=0.02$ at small and large $P$. Fig.\ref{data}(c)
shows that state at $E=0.5$ is always localized at small $P$ and
extended only for large $P(>1)$ where all curves of different
system sizes tend to merge together.

It is well-known that a fundamental difference between $P(s)$ of
localized and extended states is its behavior at small $s$. When
$s$ tends to zero, $P(s)$ tends to zero for extended states due to
level-repulsion while for localized states it tends to one due to
level-aggregation\cite{mehta}. Thus we need to consider the
behavior of $P(s)$ at small $s$ for further test of the results in
the last paragraph. It is convenient to consider a function of
integrated level-spacing distribution at small $s$ defined by
$I_P(s)=\int_{0}^{s}P(s^{\prime})ds^{\prime}$. The meaning of
$I_P(s)$ is the integrated fraction of level-spacings smaller than
$s$. Although $P(s)$ in most cases of our numerical results is
close to the GUE distribution, level-repulsion of extended states
and level-aggregation of localized states should still be expected
at small $s$. This leads to the following criteria for
localization property : $I_P(s)$ at small $s$ should increase with
increasing $L$ for localized states while decrease or remain
unchanged with increasing $L$ for extended states. Thus the
behavior of $I_P(s)$ at small $s$ can serve as another method to
distinguish between extended and localized states.
Fig.\ref{ps_small1} shows $I_P(s)$ for $(E=0,P=0.1)$ (a),
$(E=0.02,P=0.1)$ (b) and $(E=0.5,P=1.5)$ (c) for $L=8, 12, 16, 20,
24$ and comparison with $I_P(s)$ of $P_{GUE}(s)$.
Fig.\ref{ps_small2} is for $(E=0,P=0.7)$ (a), $(E=0.02,P=0.7)$ (b)
and $(E=0.5,P=0.5)$ (c). One can see clearly that states in
Fig.\ref{ps_small1} show the feature of extended states while
states in Fig.\ref{ps_small2} are localized. In order to exam each
electronic state of fixed electronic energy $E$ in the whole range
of mixing, we consider $I_P(s=0.5)$, the fraction of the
level-spacings less than $0.5$. We plot the results of $I_P(s=0.5)$
vs. $P$ at $E=0,0.02$ and $0.5$ for L=8, 12, 16, 20, 24
in Fig.\ref{ps_small3}. Similar with the criteria for $I_P(s)$, we
use the following criteria. If $I_P(0.5)$ of a state increases
with increasing $L$, the state is localized. Otherwise, they are
extended. According to this criteria, curves in
Fig.\ref{ps_small3} lead to essentially the same results for
localization property as obtained by the analysis of $I_0$ in
Fig.\ref{data}, consistent with the results in the last paragraph.

Let us now turn to the region of large $s$. Since $P_{GUE}(s)$
decays faster than $P_{PE}(s)$ at large $s$, the behavior of
$P(s)$ in this region can also be used to show difference between
extended states and localized states. At this region it is
convenient to consider another function of integrated
level-spacing distribution defined by
$F(s)=\int_{s}^{\infty}P(s)ds=1-I_P(s)$. The meaning of $F(s)$ is
the integrated fraction of level-spacings larger than $s$. Since
$F(s)$ of $P_{GUE}(s)$ is less than that of $P_{PE}(s)$ at large
$s$, we may expect that $F(s)$ at larger $s$ decreases or remains
unchanged with increasing $L$ for extended states while it
increases with increasing $L$ for localized states.
Fig.\ref{ps_tail1} shows curves of $F(s)$ at $(E=0,P=0.1)$ (a),
$(E=0.02,P=0.1)$ (b), and $(E=0.5,P=1.5)$ (c) for $L=8, 12, 16,
20, 24$ and comparison with that of $P_{GUE}(s)$.
Fig.\ref{ps_tail2} shows $(E=0,P=0.7)$ (a), $(E=0.02,P=0.7)$ (b),
and $(E=0.5,P=0.5)$ (c). In view of Fig.\ref{ps_small1} and
Fig.\ref{ps_small2}, it is clear that the results of $F(s)$
coincide with those of $I_P(s)$ concerning the localization
property. We also calculate $F(s=2)$ for each fixed energy in the
whole range of mixing. As shown above, the same criteria as that
for $I_0$ and $I_P(s=0.5)$ can be employed. The curves of $F(s=2)$
vs. $P$ are plotted in Fig.\ref{ps_tail3} for $E=0$ (a), $E=0.02$
(b) and $E=0.5$ (c). One can see that they are consistent with the
results of $I_0$ (Fig.\ref{data}) and $I_P(s=0.5)$
(Fig.\ref{ps_small3}).

\subsection{Discussion of the localization property}

In the last subsection, we analyzed the global shape of $P(s)$ and
its behavior at small and large $s$ by considering $I_0$, $I_P(s)$
and $F(s)$, respectively. Analysis of all these quantities leads
to essentially the same conclusion concerning the localization
property, as follows. For zero interband mixing, only states at
the two LB centers are extended. In the presence of interband
mixing, new extended states emerge. States near the LB centers,---
i.e., $E\sim0$,--- are delocalized by weak interband mixing and
localized by strong mixing, with a transition point at some
intermediate mixing $P_c$. For states near the region between two
LBs,---i.e., $E\sim0.5$,--- they are localized at both weak and
intermediate mixing and delocalized by strong mixing.

The existence of new extended states at $E\sim0.5$ in the case of
strong interband mixing can be understood as follows. Assume that
the intra-band tunneling at nodes are negligibly weak for states
of $E\sim0.5$, we saw already from Fig. \ref{network}(a) that the
maximum interband mixing ($\sin\theta=1$) delocalizes the state,
which is localized at zero interband mixing. If one views
$p=\sin^2\theta$ as connection probability of two neighboring
loops of opposite chirality, our two-channel model without
intra-band tunnelings at nodes is analogous to a bond-percolation
problem. It is well-known that a percolation cluster exists at
$p\ge p_c=1/2$ or $P\ge P_c=1$ for a square lattice\cite{stauff}.
Therefore, an extended state is formed by strong mixing. One hopes
that the intra-band tunnelings at nodes will only modify the
threshold value of the mixing strength.
\begin{figure}[ht]
\begin{center}
  \includegraphics[width=8cm,height=4.9cm]{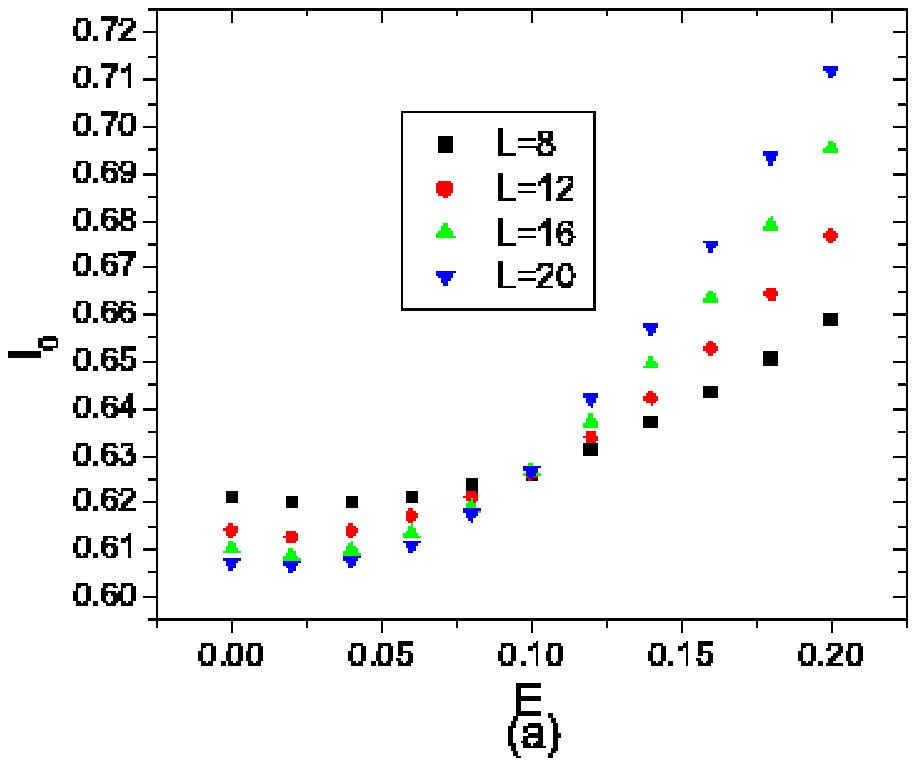}
\end{center}
\begin{center}
  \includegraphics[width=8cm,height=4.9cm]{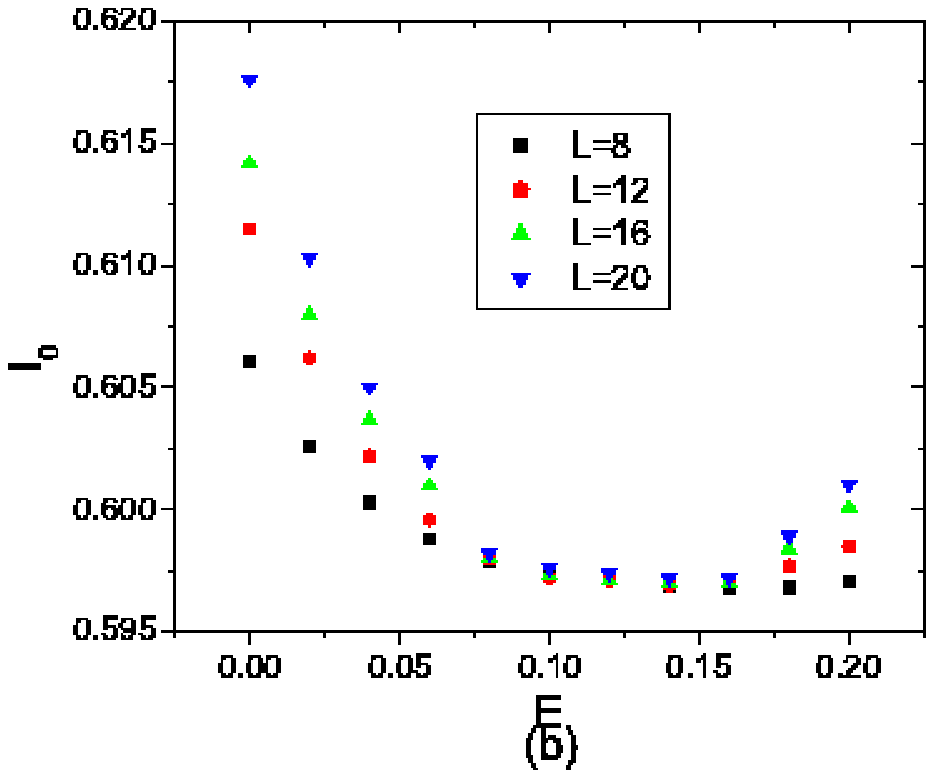}
\end{center}
\begin{center}
  \includegraphics[width=8cm,height=4.9cm]{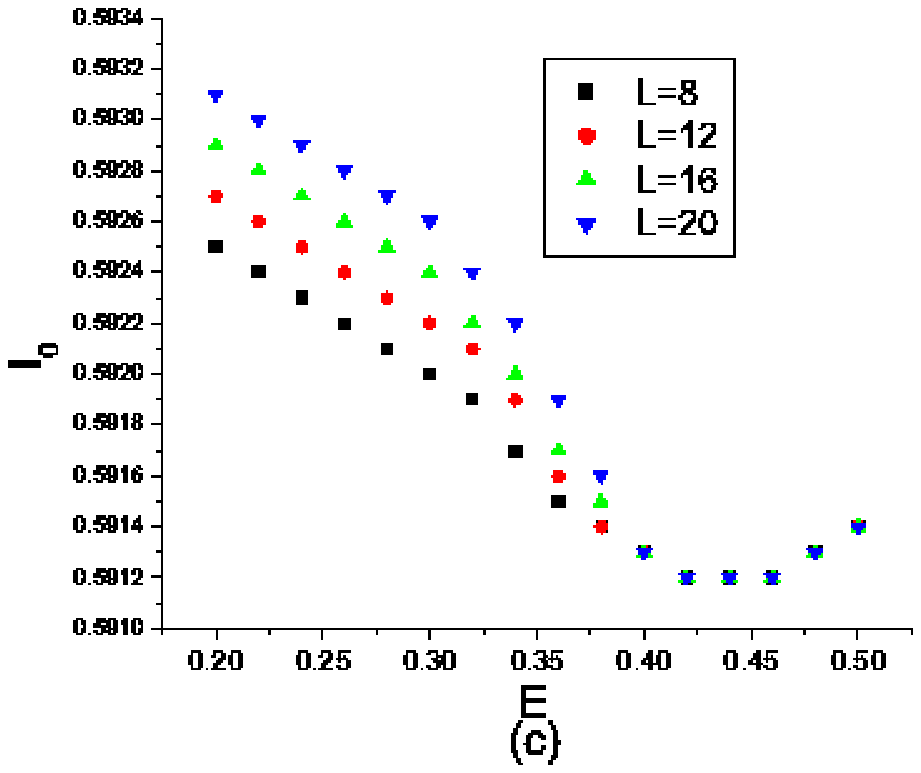}
\end{center}
\caption{$I_0$ vs. $E$ for $L=8, 12, 16, 20$, (a) $P=0.1$; (b)
$P=0.7$; (c) $P=1.5$.} \label{fixed_P}
\end{figure}

In order to show explicitly the existence of a narrow band of
extended states and its evolution with increasing mixing, curves
of $I_0$ vs. $E$ are plotted for three values of $P$ in
Fig.\ref{fixed_P}. A band of extended states is formed around the
LB center $E\in[0,0.1]$ for $P=0.1$. When $P$ is increased to an
intermediate value $0.7$, this band of extended states is lifted
up to $E\in[0.8,1.6]$. For strong mixing, it is further shifted to
$E\in[0.4,0.5]$. Thus the band of extended states in the lower LB
emerges in weak mixing and tends to float up in energy with
increased mixing. By symmetry one can expect that the extended
band of the upper LB should tend to dive down in energy with
increased mixing. The two bands of extended states in the lower
and upper LBs should finally meet at the middle energy region in
the case of strong mixing.

The above results are restricted to the case of two LBs, while
there can be infinite LBs in the continuum model for a realistic
system. Combine our above results and the float-up-merge picture
proposed by Sheng et. al.\cite{sheng2}, one can expect the
following conclusions for multiple LBs. In the case of weak
interband mixing, a narrow extended band emerges in each LB. With
increasing mixing, i.e., increasing disorders or decreasing
magnetic field, the extended band in the lowest LB floats up and
finally merges with that in the second lowest LB. Then, this
extended band will further shift up and merge into that in the
third lowest LB, and so on so forth.

\begin{figure}[ht]
\begin{center}
  \includegraphics[width=8cm,height=4.2cm]{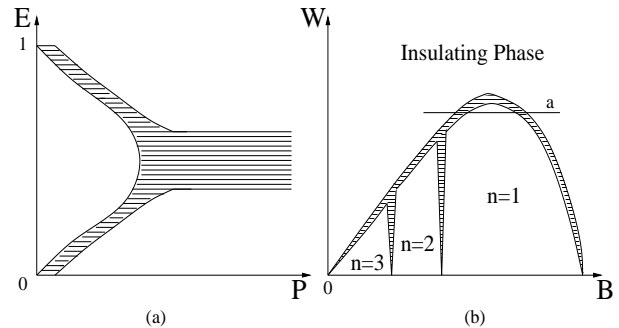}
\end{center}
\caption{(a) Topological phase diagram of electron localization in
$E-P$ plane. The shadowed regime is for extended states (metallic
phase). (b) Topological QH phase diagram in $W-B$ plane. $W$
stands for the disorder strength, and $B$ for the magnetic field.
The shadowed regime is for the metallic phase. The area indicated
by the symbol $n$ is the $n$-plateau IQHE phase. The rest area is
the insulating phase.} \label{phase}
\end{figure}

To express our numerical results in the plane of energy and
interband mixing, a topological phase diagram shown in Fig.
\ref{phase}(a) is obtained. In the absence of interband mixing,
only the singular energy level at each LB center is extended. In
the presence of interband mixing of opposite chirality, there are
two regimes. At weak mixing, each of the extended states broadens
into a narrow band of extended states near the LB centers. With
increased mixing, the extended states in the lowest LB shift from
the LB center(see Fig.\ref{fixed_P}). These extended states will
eventually merge with those from the higher LBs. This shifting of
extended states was also observed before\cite{kivelson}. At strong
mixing, a band of extended states exists between neighboring LBs
where all states are localized without the mixing.

Let us look at the consequences of the above results. For weak
disordered systems in IQHE regime, the Landau gap is larger than
the LB bandwidth. Thus there is no overlap between adjacent LBs.
According to the semiclassical picture, electronic states between
the two adjacent LBs should be from either the upper or the lower
bands with the same chirality in this case. It means that no
interband mixing occurs and there is only one extended state in
each LB. This may explain why scaling behaviors were observed for
plateau transitions in early experiments on clean samples.
Interband mixings occur when the Landau gap is less than the LB
bandwidth. Systems of relatively strong disorders in IQHE regime
should correspond to this case. As the single extended state at
each LB center broadens into a narrow extended band, a narrow
metallic phase emerges between two neighboring IQHE phases. Thus
each plateau transition contains two consecutive quantum phase
transitions for strongly disordered systems. The bands of extended
states will merge together in strong mixing. This strong mixing
regime corresponds to the case when the Landau gap is much smaller
than the bandwidth. Since the Landau gap is proportional to the
magnetic field, the disordered system should always enter the
strong mixing regime before it reaches the weak field insulating
phase, regardless of how weak the disorders are. In terms of QH
plateau transitions, a direct transition occurs because a narrow
metallic phase exists between two QH phases in a weak field. Thus,
we propose that a direct transition from an IQHE phase to the
insulating phase at weak field is realized by passing through a
metallic phase, and it should hold for both weak and strong
disordered systems.

\begin{figure}
\begin{center}
\includegraphics[width=8cm,height=5cm]{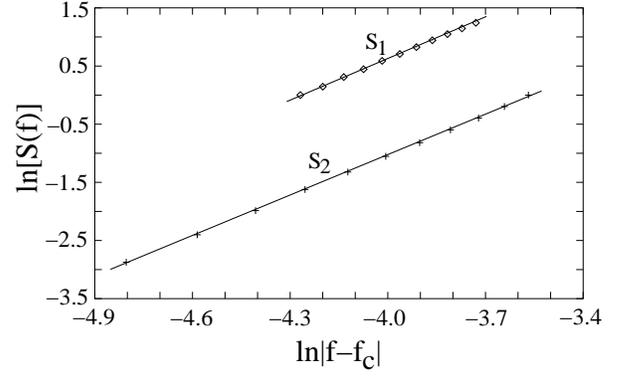}
\end{center}
\caption{The fitting result of two critical points at the left and
the right side. The two straight lines show coincidence with the
scaling law. The critical filling factors are $f_{c1}=0.6453$ and
$f_{c2}=0.6477$. The two critical exponents are equal to the value
$z\nu=2.33\pm0.01$.} \label{twopoint}
\end{figure}

\begin{figure}
\begin{center}
\includegraphics[width=8cm,height=5cm]{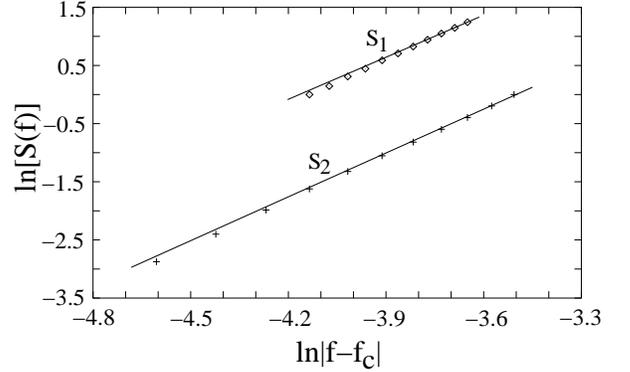}
\end{center}
\caption{The best fitting result of one single critical point at
the left and right side. The critical filling factor is
$f_{c}=0.646$. The two straight lines illustrate systematic
deviations from the scaling law at regions close to the critical
point. The average values of the two critical exponents are
$z_1\nu_1=2.58\pm0.02$ and $z_2\nu_2=2.60\pm0.02$, respectively.}
\label{onepoint}
\end{figure}

Plot above results in the plane of disorder and the magnetic
field, we obtain a new topological QH phase diagram as shown in
Fig. \ref{phase}(b). This is similar to the empirical diagram
obtained experimentally in Ref. 14. The origin ($W=0, \ B=0$) is a
singular point. According to the weak localization
theory\cite{abrahams}, no extended state exists at this point.
Differing from existing theories, there exists a narrow metallic
phase between two adjacent IQHE phases and between an IQHE phase
and an insulating phase. This new phase diagram is consistent with
the non-scaling experiments\cite{hilke} where samples are
relatively dirty, and interband mixing is strong, corresponding to
a process along line $a$ in Fig. \ref{phase}(b). The system
undergoes two quantum phase transitions each time it moves from
the QH insulating phase to IQHE phase of $n=1$ and back to the
weak field insulating phase as the magnetic field decreases.
\begin{figure}[ht]
\begin{center}
\includegraphics[width=8cm,height=5cm]{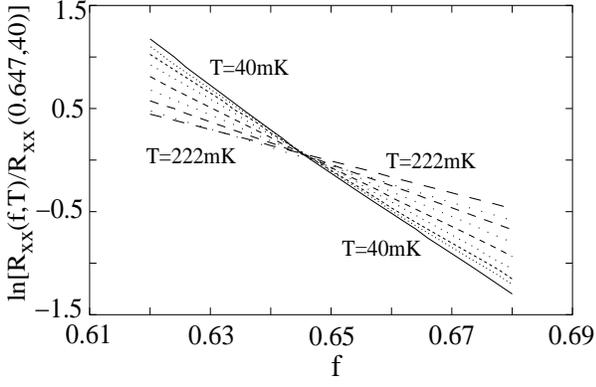}
\end{center}
\caption{\label{experiment} Experiment data of the logarithm of
the longitudinal resistance $\ln[R_{xx}(f,T)/R_{xx}(0.647,40mK)]$
in Ref. 17.
$f$ is the filling factor of LBs and $T$ is the temperature.}
\end{figure}
To verify this claim, we analyzed the original experimental data in
Ref. 17 according to the assumption of two quantum 
phase transition points. The experiment data of the logarithm of
the longitudinal resistance $ln[R_{xx}(f,T)]$ are shown in Fig.
\ref{experiment} where $f$ is the filling factor of LBs and $T$
is the temperature. According to the theory of continuous
transitions, one should obtain
\begin{equation}
    ln[R_{xx}(\nu,T)]=F_1(S_1(f)/T)
\end{equation}
with $S_1(\nu)\sim(f_{c1}-f)^{z_1\nu_1}$ for the region of
$f<f_{c1}$ while
\begin{equation}
    ln[R_{xx}(\nu,T)]=F_2(S_2(f)/T)
\end{equation}
with $S_2(\nu)\sim(f-f_{c2})^{z_2\nu_2}$ for the region of
$f>f_{c2}$. Previous theories predict one single critical point,
---i.e., $f_{c1}=f_{c2}$ and $z_1\nu_1=z_2\nu_2$. But our results
suggest two distinct critical points. By standard scaling
analysis, two good scaling behaviors are obtained for two close
critical filling factors of $f_{c1}=0.6453$ and $f_{c2}=0.6477$ as
shown in Fig. \ref{twopoint}. The critical exponents in both the
left side and the right side of the transition region are equal to
the value $z\nu=2.33\pm0.01$. On the other hand, the fit for one
single critical point fails. Fig. \ref{onepoint} shows the result
of a single critical point at $\nu_c=0.646$. It is the best
fitting result for a single critical point if we require that the
two critical exponents are approximately equal and the scaling law
is optimally obeyed. The two critical exponents are
$z_1\nu_1=2.58\pm0.02$ and $z_2 \nu_2 = 2.60 \pm 0.02$, deviating
from the theoretical results $\nu \sim 2.33$. One can also see
clearly systematic deviations from the scaling law in the region
close to the critical point at both sides in Fig. \ref{onepoint}.
This implies that the transition process is governed by two
separated critical points instead of one. The regime between the
two critical points should correspond to the metallic phase. Our
fitting shows that the width of this regime is about
$5\times10^{-3} tesla$ while the value of the magnetic field was
increased by $1\sim2\times10^{-3} tesla$ each time in the
experiments. This may explain why the metallic phase was
overlooked.

It is worth noting that there is another puzzle in the non-scaling
experiment which may be solved by our two-critical-point picture.
To be specific, let us consider the experimental data for the
transition between the QH insulating phase and the $n=1$ IQHE
phase. It was shown that the logarithm of the longitudinal
resistance $ln[R_{xx}(f,T)]$ can be fitted with a linear function
of the LB filling factor $f$ (see Fig.\ref{experiment}) as
following\cite{hilke}
\begin{equation}
  ln[R_{xx}(f,T)]=ln[R_{xx}(f_c,T)]-(f-f_c)/(\alpha+\beta T)]
\end{equation}
where $\alpha$ and $\beta$ are positive constants, $f_c$ is the
filling factor where curves of different temperature $T$ cross
approximately. Since $\alpha$ is non-zero\cite{hilke}, it leads to
the conclusion that $R_{xx}(f,T)$ at the limit of $T=0$ remains
finite {\it for every $f$}. This is puzzling because it is
inconsistent with the theoretical requirement that
$R_{xx}(T=0)=\infty$ in the QH insulating phase, i.e., $f<f_c$,
and $R_{xx}(T=0)=0$ in the $n=1$ IQHE phase, i.e., $f>f_c$. This
puzzle may be solved as follows. Combine the linear relationship
between $ln[R_{xx}(f,T)]$ and $f$ for fixed $T$ with our picture
of two critical points $f_{c1}<f_{c2}$, we expect
\begin{equation}
  ln[R_{xx}(f,T)]=ln[R_{xx}(f_{c1},T)]-(f-f_{c1})/(A_1 T^{z\nu})
\end{equation}
in the QH insulating phase, i.e., $f<f_{c1}$, while
\begin{equation}
  ln[R_{xx}(f,T)]=ln[R_{xx}(f_{c2},T)]-(f-f_{c2})/(A_2 T^{z\nu})
\end{equation}
in the n=1 IQHE phase, i.e., $f>f_{c2}$, where $A_1$ and $A_2$ are
positive constants, and $z$ and $\nu$ are critical exponents. It
is clear that both $R_{xx}(f,T=0)=\infty$ in $f<f_{c1}$ and
$R_{xx}(f,T=0)=0$ in $f>f_{c2}$ are recovered. While a finite
value of $R_{xx}(f,T=0)$ in the region $f_{c1}<f<f_{c2}$ is
consistent with our prediction of a metallic phase between the two
critical points.

Our model can also be used to describe spin-polarized systems. In
this case, the two LBs are for spin up and spin down states.
Indeed, two-channel CC models have been used before to simulate a
spin-resolved problem\cite{wang}. In the presence of interband
mixing, two distinct critical points were obtained. They were
related to the two extended states in the two subbands, which are
shifted by the mixings. However, this study could not discern
whether the states in between are extended or localized. Thus it
was not clear whether the extended states in the presence of
mixing are just the two points or form a band. In this sense, our
results are consistent with those of early works. It is worth
noting that the spin resolved problem is different from the
non-scaling experiment. In the spin resolved problem, the energy
region considered includes the centers of both the spin-up and
spin-down subbands. Thus there are two near-degenerate extended
states in this region in the absence of interband mixing. It is
then natural to regard the two distinct critical points as two
distinct extended states of the two subbands in the presence of
mixing. However, in the non-scaling experiment, the region
considered includes only the center of the lowest LB. Thus there
is only one extended state in the absence of interband mixing. In
this case, the two separate critical points may not be considered
as two distinct extended states. It seems that the only suitable
explanation is the existence of a band of extended states between
the two points.

One should also notice that two types of metallic phases have been
studied extensively in the QH system. One is the composite Fermion
state at the half-filling in the lowest Landau level (LL) and the
other is the stripe state at the half-filled higher LLs. These
states are formed by the Coulomb interaction effect in the high
mobility samples. They are different from our metallic phase due
to level mixing in the paper. Although we have not considered the
electron-electron interactions in our study, there is no reason
why the delocalization effect dof level mixing will be diminished
by the Coulomb interaction. Of course, the interaction could
change the level mixing effect's dependence on the magnetic field.

\section{Conclusions}
In conclusion, we find by numerical calculations within the
network model that the single extended state at each LB center in
the absence of interband mixing broadens into a narrow band of
extended states when the effect of mixing of states of {\it
opposite chirality} is taken into account. With the decrease of
magnetic field or increase of disorders, these extended bands
further broaden and may merge together. Based on this, we propose
a new phase diagram in which a narrow metallic phase exists
between two neighboring IQHE phases and between an IQHE phase and
an insulating phase. This new phase diagram is consistent with
non-scaling behaviors observed in recent experiments. A standard
scaling analysis on experiment data in Ref. 17 
supports our results.

\begin{acknowledgments}
This work was substantially supported by a grant from the Research
Grant Council of HKSAR, China (Project No. HKUST6153/99P, and
HKUST6149/00P). GX gratefully acknowledges the support of National
Administration Foundation of P. R. China and the K. C. Wong's
Education Foundation, Hong Kong. GX and YW also acknowledge the
support of CNSF under grant No. 90103024.
\end{acknowledgments}

\appendix*
\section{}
\begin{figure}[ht]
\begin{center}
  \includegraphics[height=7cm, width=8cm]{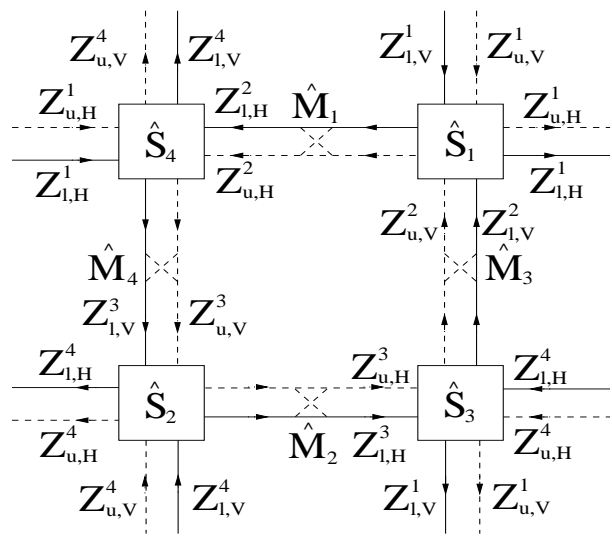}
\end{center}
\caption{A two-channel network model of $2\times2$ nodes with
periodic boundaries along both directions. $Z$'s are the
wavefunction amplitudes on links. The notations are as follows. H
and V stand  for horizontal and vertical links, respectively. $u$
($l$) is for the upper (lower) LB. $\hat{S}_i$ are SO(4) matrices
describing tunneling at nodes, and $\hat{M}_i$ are $U(2)$ matrices
for interband mixing.} \label{example}
\end{figure}

In this appendix, we explicitly construct the evolution matrix
$\hat U$ for a $2\times 2$ two-channel CC-network model as shown
in Fig. \ref{example}. Periodical boundary conditions in both
directions are imposed as explained in section III. $Z$-s are the
wavefunction amplitudes on links. The notations are as follows. H
and V stand  for horizontal and vertical links, respectively. $u$
($l$) is for the upper (lower) LB. $\hat{S}_i$ are SO(4) matrices
defined in Eq. \ref{smatrix} describing the tunneling on nodes,
and $\hat{M}_i$ are $U(2)$ matrices defined in Eqs. \ref{mmatrix1}
and \ref{mmatrix2} describing interband mixing. From Fig.
\ref{example} we can obtain
\begin{equation}
\left (
\begin{array}{l}
Z_{u,H}^{1}(t+1) \\ Z_{l,H}^{1}(t+1) \\
Z_{l,H}^{2}(t+1) \\ Z_{u,H}^{2}(t+1)
\end{array}
\right ) =
\hat{H}_1
\left (
\begin{array}{l}
Z_{l,V}^{1}(t)\\ Z_{u,V}^{1}(t)\\ Z_{u,V}^{2}(t) \\
Z_{l,V}^{2}(t)
\end{array}
\right )
\end{equation}
\begin{equation}
\left (
\begin{array}{l}
Z_{u,H}^{3}(t+1) \\ Z_{l,H}^{3}(t+1) \\
Z_{l,H}^{4}(t+1) \\ Z_{u,H}^{4}(t+1)
\end{array}
\right ) =
\hat{H}_2
\left (
\begin{array}{l}
Z_{l,V}^{3}(t) \\ Z_{u,V}^{3}(t) \\ Z_{u,V}^{4}(t)
\\ Z_{l,V}^{4}(t)
\end{array}
\right )
\end{equation}
\begin{equation}
\left (
\begin{array}{l}
Z_{l,V}^{1}(t+1) \\ Z_{u,V}^{1}(t+1) \\
Z_{u,V}^{2}(t+1) \\ Z_{l,V}^{2}(t+1)
\end{array}
\right ) =
\hat{H}_3
\left (
\begin{array}{l}
Z_{u,H}^{3}(t) \\ Z_{l,H}^{3}(t) \\ Z_{l,H}^{4}(t)
\\ Z_{u,H}^{4}(t)
\end{array}
\right )
\end{equation}
\begin{equation}
\left (
\begin{array}{l}
Z_{l,V}^{3}(t+1) \\ Z_{u,V}^{3}(t+1) \\
Z_{u,V}^{4}(t+1) \\ Z_{l,V}^{4}(t+1)
\end{array}
\right ) =
\hat{H}_4
\left (
\begin{array}{l}
Z_{u,H}^{1}(t) \\ Z_{v,H}^{1}(t) \\ Z_{l,H}^{2}(t)
\\ Z_{u,H}^{2}(t)
\end{array}
\right ),
\end{equation}
with
\begin{equation}
   \hat{H}_1=
\left (
\begin{array}{ll}
\hat{1} & \hat{0} \\
\hat{0} & \hat{M}_1
\end{array}
\right )
\hat{S}_1; \quad
   \hat{H}_2=
\left (
\begin{array}{ll}
\hat{M}_2 & \hat{0} \\
\hat{0} & \hat{1}
\end{array}
\right )
\hat{S}_2; \nonumber
\end{equation}
\begin{equation}
   \hat{H}_3=
\left (
\begin{array}{ll}
\hat{1} & \hat{0} \\
\hat{0} & \hat{M}_3
\end{array}
\right )
\hat{S}_3; \quad
   \hat{H}_4=
\left (
\begin{array}{ll}
\hat{M}_4 & \hat{0} \\
\hat{0} & \hat{1}
\end{array}
\right )
\hat{S}_4,  \nonumber
\end{equation}
where $\hat{1}$ and $\hat{0}$ are the $2\times 2$ identity
and zero matrices, respectively.
If we define
\begin{eqnarray}
\phi_{H} = \left (
        \begin{array}{l}
Z_{u,H}^{1}\\ Z_{l,H}^{1}\\
Z_{l,H}^{2}\\ Z_{u,H}^{2}\\
Z_{u,H}^{3}\\ Z_{l,H}^{3}\\
Z_{l,H}^{4}\\ Z_{u,H}^{4}\\
        \end{array}
        \right ); \quad
\phi_{V} = \left (
        \begin{array}{l}
Z_{l,V}^{1}\\ Z_{u,V}^{1}\\
Z_{u,V}^{2}\\ Z_{l,V}^{2}\\
Z_{l,V}^{3}\\ Z_{u,V}^{3} \\
Z_{u,V}^{4}\\ Z_{l,V}^{4} \\
        \end{array}
        \right ), \nonumber
\end{eqnarray}
then the evolution equation is
\begin{equation}
        \left (
        \begin{array}{l}
        \phi_H(t+1) \\ \phi_V(t+1)
        \end{array}
        \right )
        =
        \hat{U}
        \left (
        \begin{array}{l}
        \phi_H(t) \\ \phi_V(t)
        \end{array}
        \right ).
\end{equation}
The evolution operator $\hat{U}$ is
\begin{equation}
        \hat{U}
        =
        \left (
        \begin{array}{llll}
        \hat{0} & \hat{0} & \hat{0} & \hat{H}_1 \\
        \hat{0} & \hat{0} & \hat{H}_2 & \hat{0} \\
        \hat{H}_3 & \hat{0} & \hat{0} & \hat{0} \\
        \hat{0} & \hat{H}_4 & \hat{0} & \hat{0}
        \end{array}
        \right ),
\end{equation}
where $\hat{0}$ is the $4\times 4$ zero matrix. It has the
structure of Eq. \ref{evolution}.

\end{document}